\documentclass[twocolumn]{aastex631}

\usepackage{CJK}
\usepackage{amsfonts}
\usepackage{amssymb}
\usepackage{subfigure}
\usepackage{graphicx}
\usepackage{ulem}
\usepackage{amsmath}
\usepackage{color}
\usepackage{xcolor}
\usepackage{float}
\usepackage{makecell}
\usepackage{newtxtext,newtxmath}
\usepackage{mathrsfs}
\usepackage{gensymb}
\usepackage{multirow}
\usepackage{booktabs}
\usepackage{threeparttable}
\usepackage{ulem}
\usepackage{rotating}
\usepackage{hyperref}
\usepackage{graphicx}

\newcommand{\mr}{\rm}
\newcommand{\MHz}{\ensuremath{\, {\rm {MHz}}}}

\newcommand{\km}{\ensuremath{\, {\rm {km}}}}
\newcommand{\m}{\ensuremath{\, {\rm {m}}}}
\newcommand{\cm}{\ensuremath{\, {\rm {cm}}}}
\newcommand{\K}{\ensuremath{\, {\rm {K}}}}

\shorttitle{21 cm global spectrum measurement on lunar orbit} 
\shortauthors{Shi et al.}

\begin{document}
\begin{CJK*}{UTF8}{gbsn}

\title{Lunar Orbit Measurement of Cosmic Dawn 21 cm Global Spectrum }
\author[0000-0001-8233-3703]{Yuan Shi (施\hbox{\scalebox{0.4}[1]{女}\kern-.13em\scalebox{0.7}[1]{原}})}
\affiliation{National Astronomical Observatories, Chinese Academy of Sciences, Beijing 100101, China}
\affiliation{School of Astronomy and Space Science, University of Chinese Academy of Sciences, Beijing 100049, China}

\author{Furen Deng (邓辅仁)}
\affiliation{National Astronomical Observatories, Chinese Academy of Sciences, Beijing 100101, China}
\affiliation{School of Astronomy and Space Science, University of Chinese Academy of Sciences, Beijing 100049, China}

\author[0000-0003-3224-4125]{Yidong Xu (徐怡冬)}
\affiliation{National Astronomical Observatories, Chinese Academy of Sciences, Beijing 100101, China}

\author[0000-0002-6174-8640]{Fengquan Wu (吴锋泉)}
\affiliation{National Astronomical Observatories, Chinese Academy of Sciences, Beijing 100101, China}

\author{Qisen Yan (严琦森)}
\affiliation{School of Mechanical Engineering, Hangzhou Dianzi University, Hangzhou 310018, China}
\affiliation{National Astronomical Observatories, Chinese Academy of Sciences, Beijing 100101, China}

\author[0000-0001-6475-8863]{Xuelei Chen (陈学雷)}
\affiliation{National Astronomical Observatories, Chinese Academy of Sciences, Beijing 100101, China}
\affiliation{School of Astronomy and Space Science, University of Chinese Academy of Sciences, Beijing 100049, China}
\affiliation{Department of Physics, College of Sciences, Northeastern University, Shenyang 110819, China}
\affiliation{Center for High Energy Physics, Peking University, Beijing 100871, China}

\correspondingauthor{Yidong Xu, Xuelei Chen}
\email{xuyd@nao.cas.cn, xuelei@cosmology.bao.ac.cn}


\begin{abstract}
A redshifted 21 cm line absorption signature is commonly expected from the cosmic dawn era, when the first stars and galaxies formed. The detailed traits of this signal can provide important insight on the cosmic history. However, high precision measurement of this signal is hampered by the ionosphere refraction and absorption, as well as radio frequency interference (RFI). A space observation can solve the problem of the ionosphere, and the Moon can shield the RFI from the Earth. In this paper, we present simulations of the global spectrum measurement in the 30 -- 120 MHz frequency band on the lunar orbit, from the proposed Discovering the Sky at the Longest wavelength (DSL) project. In particular, we consider how the measured signal varies as the satellite moves along the orbit, take into account the blockage of different parts of the sky by the Moon and the antenna response. We estimate the sensitivity for such a 21 cm global spectrum experiment. An RMS noise level of $\leq 0.05 \K$ is expected at 75 MHz after 10 orbits ($\sim$ 1 day) observation, for a frequency channel width of 0.4 MHz. We also study the influence of a frequency-dependent beam, 
which may generate complex spectral structures in the spectrum. Estimates of the uncertainties in the foreground and 21 cm model parameters are obtained.
\end{abstract}

\keywords{Observational cosmology(1146) --- H I line emission(690) --- Reionization(1383) --- Radio telescopes(1360)
--- Space telescopes(1547)}


\section{Introduction}          
\label{sect:intro}
Depending on whether the spin temperature of the neutral hydrogen atoms is higher or lower than that of the cosmic background radiation, there is net emission or absorption of radiation of 21 cm wavelength. Such emission or absorption features from the early universe provide us a useful tool to study the cosmic dark ages, the cosmic dawn (CD) and the epoch of reionization (EoR). 

The 21~cm global spectrum experiments aim to measure the sky-averaged spectrum with high precision, so as to probe the early epochs of the Universe. There are a number of such ground-based experiments, including the the Experiment to Detect the Global Epoch-of-Reionization Signature (EDGES, \citealt{2008ApJ...676....1B, 2010Natur.468..796B, 2017ApJ...835...49M}), the Sonda Cosmológica de las Islas para la Detecciónde Hidrógeno Neutro (SCI-HI, \citealt{2014ApJ...782L...9V}), the Probing Radio Intensity at high-z from Marion (PRIzM, \citealt{2019JAI.....850004P}), the Shaped Antenna measurement of the background RAdio Spectrum (SARAS, \citealt{2013ExA....36..319P, 2017ApJ...845L..12S, 2018ApJ...858...54S}), the Cosmic Twilight Polarimeter (CTP, \citealt{2019ApJ...883..126N}), the Broadband Instrument for Global Hydrogen Reionization Signal(BIGHORNS; \citealt{2015PASA...32....4S}), the Large-Aperture Experiment to Detect the Dark Age (LEDA, \citealt{2016MNRAS.461.2847B, Bernardi2018:1802.07532v1, 2018MNRAS.478.4193P}), and the Radio Experiment for the Analysis of Cosmic Hydrogen (REACH, \citealt{8879199}). Compared to the 21~cm tomography experiments, measurement of the global 21 cm emission has a higher raw sensitivity and requires smaller collecting area, so that it could be conducted even with a single antenna. The EDGES \citep{Bowman2018Naturea} reported the detection of a strong absorption feature at $\sim 78$~MHz which has a cosmic dawn 21 cm spectrum interpretation, though it has an unexpectedly large amplitude (0.5 K) and an unusual flattened profile. If originated from the cosmic 21 cm spectrum, this would suggest possibly new physics or astrophysics  \citep{2004ApJ...602....1C, 2008ApJ...684...18C, 2011MNRAS.415.3706C, 2013MNRAS.429.3353N, 2018Natur.555...71B, Fialkov2018PhysRevLett, 2018PhLB..785..159F, 2018PhRvD..98j3005B,2018PhRvD..98b3013S, 2021IJMPD..3050041L, 2018PhRvL.121k1301H, 2018MNRAS.480L..85H, 2018PhRvL.121l1301M, 2021PhRvD.104f3528Y}, though this result is not confirmed by a recent measurement of the SARAS experiment \citep{Singh:2021mxo}.
It is imperative to check this result and improve upon it with further and more precise observations.

In ground-based global spectrum experiments, a number of possible systematic effects, such as attenuation and differential refraction of the radio wave by the ionosphere \citep{2014MNRAS.437.1056V, 2021MNRAS.503..344S}, radio frequency interference (RFI), and the interactions of the antenna with nearby ground and underground features, can introduce distortions and errors, thus limiting the measurement precision. The magnitude of the ionospheric disruptions on global spectrum varies  in the range of $\sim 0.1\K-3 \K$ depending on the ionospheric condition.
These systematic effects can be avoided or mitigated by a space-borne experiment. A number of space mission concepts have been proposed, and some are actively studied at present, such as the Discovering the Sky at the Longest wavelength (DSL; also known as {\it Hongmeng} in Chinese, \citealt{2019arXiv190710853C,2020arXiv200715794C}), the Dark Ages Polarimetry PathfindER (DAPPER; \citealt{2018JCAP...12..015T}) and its precursor the Dark Ages Radio Explorer (DARE, \citealt{2017ApJ...844...33B}), and the Farside Array for Radio Science Investigations of the Dark ages and Exoplanets (FARSIDE, \citealt{2019arXiv191108649B,2021arXiv210308623B}). 
By orbiting the Moon, or landing on the far side of the lunar surface, the RFI from the Earth can be shielded, providing an ideal environment for such observations. 

Compared with lunar surface experiments, the lunar orbit experiment has the advantage that it is not affected by the transient ionosphere on the day-side of the lunar surface or the reflections of the lunar rocks. Engineeringly it is also simpler, as it does not require landing on the Moon, and it is easy to supply the required energy (solar power) and transmit the data back to the Earth. In the DSL mission concept, an array of satellites will be launched together as an assembly into the lunar orbit by a single rocket, then released sequentially into a linear formation on the same circular orbit. During the mission, these satellites will make both interferometric imaging \citep{Huang2018a} and high precision global spectrum measurement on the part of orbit behind the Moon, and the data will be transmitted to the Earth at the near side part of the orbit. 
\citet{Yuan2021} (hereafter referred as Paper I) have investigated the imaging quality and sensitivity of such an array in the presence of thermal noise for interferometric observation. 

For the global spectrum experiment, the lunar orbit is also an ideal option, as the ionosphere refraction and ground reflection are largely avoided. Nevertheless, as the antenna beam may have frequency dependence, it is still a complicated problem. This effect have been investigated in a number of studies. 
In the context of the EDGES experiment, \citet{2016MNRAS.455.3890M} simulated the chromatic beam effects for two dipole antennas used in observations, and assessed the detectability of the global 21 cm signal. Further, \citet{2021AJ....162...38M} validated the beam model used by EDGES by comparing different electromagnetic solvers and by comparing the simulated spectra with real observational data.
More generally, in order to extract the 21 cm signal from the foregrounds convolved by a chromatic beam,
one can 
construct training sets that include the foreground, the beam, and other instrument effects, which are then decomposed into basis vectors using Singular Value Decomposition (SVD) \citep{2018ApJ...853..187T, 2020ApJ...897..174R, 2020ApJ...897..175T, 2021ApJ...915...66T, 2021AAS...23811103B}. There are also some models to deal with systematic distortions by a Bayesian nested sampling algorithm \citep{2021MNRAS.506.2041A}.

In this paper, we explore the 21 cm global spectrum measurements on the lunar orbit. While our investigation is made in the context of the DSL mission, some of the results may also be applicable to other similar missions. 
This paper is structured as follows. Section \ref{sec:measure} introduces some basic set-ups of the global signal experiment and the tools for generating the mock spectrum: the foreground map, the 21 cm signal, and the noise and beam model. In Section \ref{sec:simulation}, we describe the simulation and the extraction of the foreground and the 21 cm signal from the mock data. 
We conclude in Section \ref{sec:conclusion}.

\begin{table}[t]
	\centering
	\caption{Basic parameters of the DSL global spectrum experiment.}
	\label{tab:basic parameters}
	\begin{tabular}{lc} 
		\hline
		Mission Parameter  & Values               \\
		\hline
		Orbit height & 300 \km\\
		Orbit radius & 2037 \km\\
		Orbital plane inclination & $30^\circ$   \\
		Precession period   & 1.3 year    \\
		Total observation time  & 3 -- 5 years \\
		High Frequency Band (HFB) & 30 -- 120 MHz \\
		Low Frequency Band (LFB) & 1 -- 30 MHz\\
		Channel bandwidth of HFB & 0.4 \MHz \\
		HFB time resolution & 8 -- 10 s\\		
		\hline
	\end{tabular}
\end{table}

\section{The global spectrum measurements}
\label{sec:measure}

The basic parameters of the DSL concept relevant to our end-to-end simulation of the global spectrum experiment are listed in Table \ref{tab:basic parameters}.
In the current DSL mission concept, a single dedicated micro-satellite would measure the global spectrum in the 30 -- 120 MHz frequency range, while a number of micro-satellites would perform both inter-satellite interferometry and single antenna global spectrum measurements below 30 MHz. The high frequency band (30 -- 120 MHz) micro-satellite uses a disc-cone antenna, while the low frequency band (1 -- 30 MHz) micro-satellites use tripole antennas. To keep a stable flying formation, the micro-satellites are tuned to have one face oriented to the Moon center, and an axis pointed along the tangent direction of the orbit, so that in the rotating frame it is kept at a stable configuration. Here as a simple model for the circular orbit, we assume the axis of the antenna cone is pointing to the center of the Moon.

With the satellites moving on the lunar orbit, parts of the sky are blocked by the Moon at any time, and the blocked region change with time. For any given instant, the visible part of the sky is determined by the position of the satellite. We treat the Moon as a solid sphere, and take a 300 km circular orbit with an inclination angle of 30$^\circ$ with respect to the lunar equator as our fiducial orbit, neglecting the variation of orbital height induced by  gravity anomaly. The orbital plane precesses with respect to the inertial frame with a cycle period of about 1.3 years, which we approximate as a uniform rotation. Simple simulation presented in paper I show that during about 31\% of the total orbital time the Earth is shielded by the Moon, and about 10\% time is ``doubly good'' during which both the Earth and the Sun are shielded.

To simulate the spectrum measurements, we generate time-ordered mock data at different frequencies assuming a sky foreground map, a 21 cm signal model and a noise model, taking into account of the satellite orbit, the Moon blockage and the antenna response. We then try to extract the 21 cm signal, and assess the measurement errors on the 21 cm model parameters.
The signal received by the antenna is modeled by the convolution of the sky signal with the antenna beam and the Moon blockage, 
\begin{equation}
T{(\nu,t)}=\int S(\mathbf{\hat{k}}, t) B(\nu, \mathbf{\hat{k}}, t) T_{\rm{sky}}(\nu, \mathbf{\hat{k}}) d^2\mathbf{\hat{k}}
\label{eq:I_intergal}
\end{equation}
where $\mathbf{\hat{k}}$ is the unit direction vector, $B(\nu, \mathbf{\hat{k}}, t)$ is the antenna primary beam response, and $S(\mathbf{\hat{k}}, t)$, which we shall call the {\it shade function},  describes the blockage of sky by the Moon. Here we treat it as a simple geometric function, neglecting frequency dependence induced by diffraction.
We pixelize the sky with the HEALPix scheme \citep{2005ApJ...622..759G}, the integral is then approximated by sum over pixels. 
\begin{equation}
T{(\nu,t)}=\sum_{k=1}^{N_{\rm pix}} S(k, t) B(\nu, k, t) T_{\rm{sky}}(\nu,k) \Delta\Omega 
\label{eq:signal}
\end{equation}
where $k$ is the pixel index, $N_{\rm pix}$ is the total number of pixels,
$\Delta\Omega=4\pi/N_{\rm pix}$ is the solid angle corresponding to one pixel. We have also checked the precision of integration by computing Eq.(\ref{eq:signal}) with $N_{\rm side}=8, 32, 64$, and 128, using the foreground model described below. The differences are smaller than $10^{-4}$ K for our sky map. Below we take HEALPIX $N_{\rm side}=64$, corresponding to a resolution of about $1^{\degree}$ which should be sufficient for the global spectrum evaluation.

\begin{figure}[ht!]
	\centering
	\includegraphics[width=0.8\columnwidth]{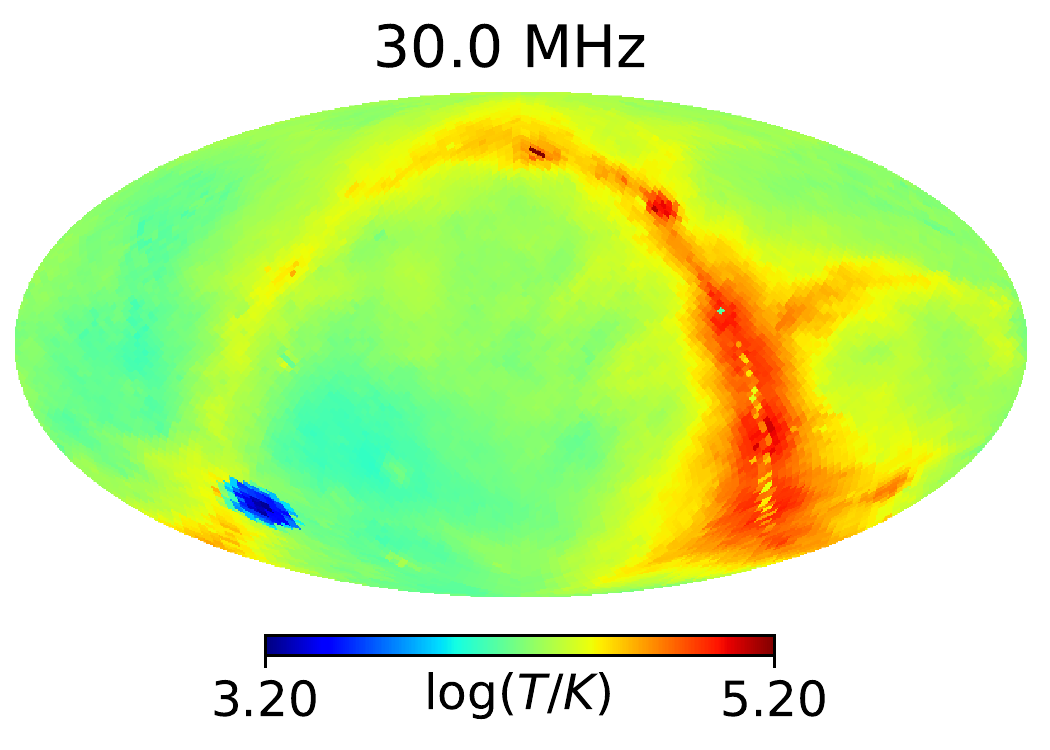}\\
	\includegraphics[width=0.8\columnwidth]{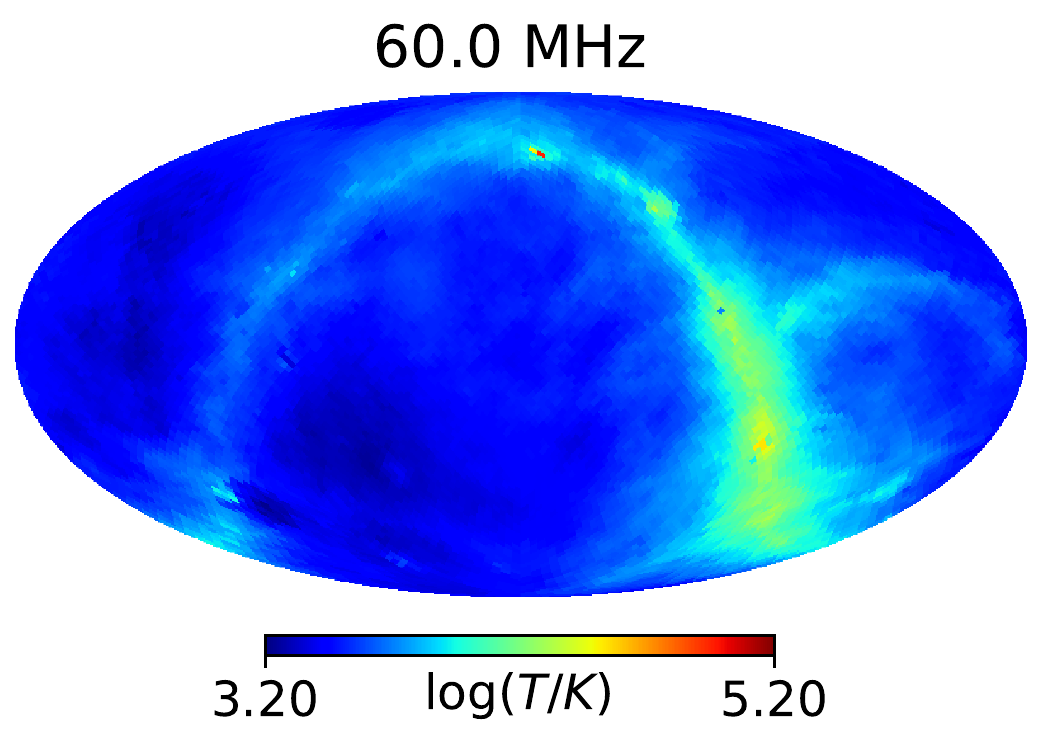}\\
    \includegraphics[width=0.8\columnwidth]{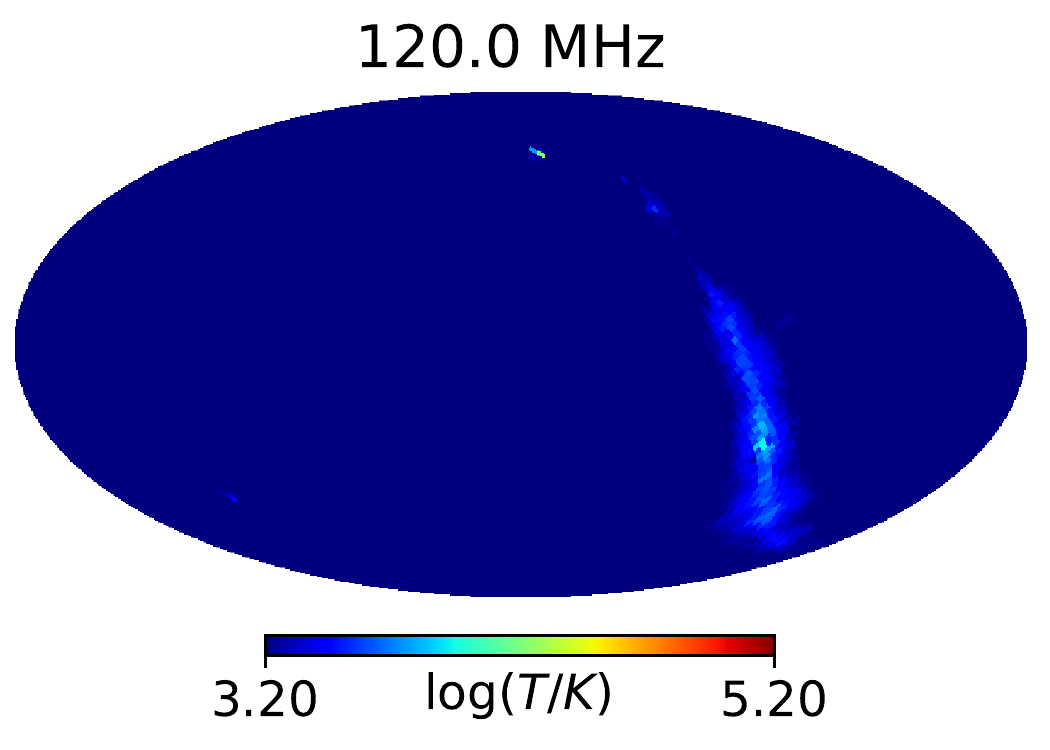}	
    \caption{The whole sky map from the ULSA model at 30 MHz, 60 MHz and 120 MHz in the ecliptic coordinate system.} 
    \label{fig:foreground_map}
\end{figure}

\subsection{Foreground model}
\label{subsec:foreground}
The mock observational data are generated with a physically-motivated sky model. The input
sky map can be obtained at a given frequency by extrapolation from observed data. A few programs are available for this purpose, e.g. the Global Sky Model (GSM) \citep{2008MNRAS.388..247D, 2017MNRAS.464.3486Z}, the Cosmology in the Radio Band (CORA) \citep{2014ApJ...781...57S}, and the Self-consistent whole Sky foreground Model (SSM) \citep{Huang2019}. However, the interstellar medium (ISM) absorption becomes quite significant at low frequencies, which may affect the foreground-21 cm signal separation. 
In this work, we make use of the high-resolution Ultra-Long wavelength Sky model with Absorption (ULSA) \citep{Cong2021ArXiv210403170Astro-Ph}, which incorporated the free-free absorption effect of the ISM\footnote{\url{https://github.com/Yanping-Cong/ULSA}}.

The sky maps at 30 MHz, 60 MHz, and 120 MHz are shown in Fig.~\ref{fig:foreground_map}. The galactic free-free absorption lowers the temperature of the sky at lower frequencies towards HII regions around massive stars or SNRs, an especially obvious one is the Gum Nebula \citep{2001MNRAS.325.1213W} at the lower left corner (around $l \sim 150^{\degree}$) at 30~MHz. In comparison to conventional absorption-free model, inclusion of the free-free absorption effect results in a decrements of 2.5\% at 30 MHz and 0.5\% at 120 MHz in sky brightness temperature.

\begin{figure}[ht!]
	\centering
	\includegraphics[width=0.9\columnwidth]{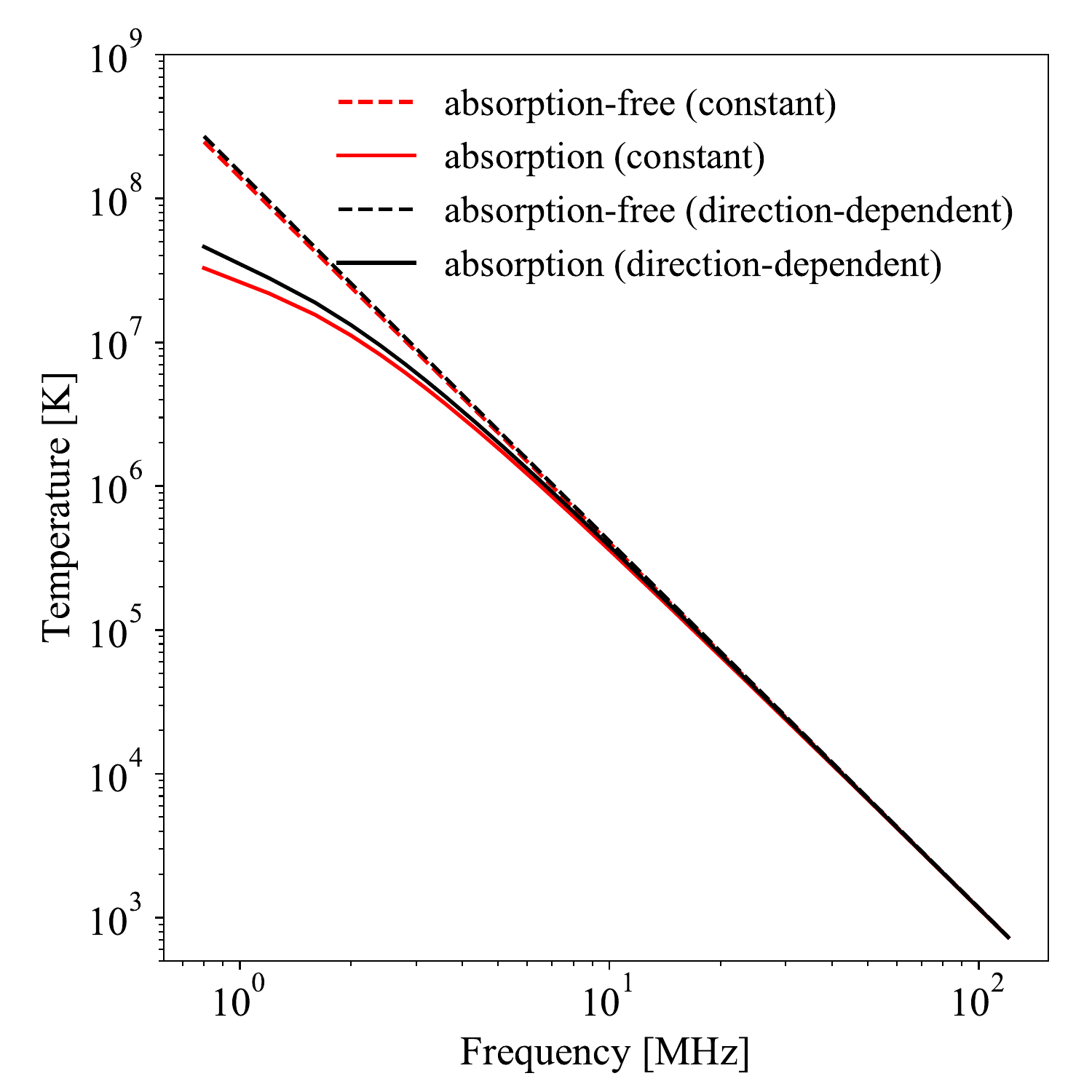}
    \caption{The input Galactic foreground global spectrum from the ULSA model. The solid lines account for the free-free absorption by the ISM, while the dashed lines are for absorption-free models. In each set of lines, the red line assumes constant spectral index, and the black line adopts a direction-dependent spectral index. 
   }
    \label{fig:foreground_signal}
\end{figure}

The ULSA model considers both cases of a constant spectral index and a direction-dependent spectral index for the emission. In the case of a direction-dependent spectral index, the spectral index map is obtained by combining the Haslam 408 MHz map \citep{Haslam1974,Haslam1981,Haslam1982,Remazeilles2015}, the LWA maps \citep{Dowell2017}, and the Guzman 45 MHz map \citep{Guzman2011} (through Eq.(29) in \citealt{Cong2021ArXiv210403170Astro-Ph}), after smoothing all maps with a Gaussian beam with FWHM $= 5\degree$.
The whole-sky averaged spectrum is shown in Fig.~\ref{fig:foreground_signal}. The solid lines are the absorption-included sky brightness, and the dashed lines are the absorption-free sky brightness. Here we show the result for both a constant and a direction-dependent spectral index of emission. We can see that there are significant differences at the lower frequencies due to absorption, which may affect the fitting of the spectrum for a given foreground model. It is therefore important to take into account of the ISM absorption if one is to fit the spectrum at the lower frequencies. Also, considering the importance of incorporating the spatial variation in the spectral index when accounting for the chromatic beam effect \citep{2021MNRAS.506.2041A},
in this work we shall use the model with absorption and  direction-dependent spectral index
to simulate the mock data.

Then in the data analysis,
the foreground contribution to the global spectrum could be modeled as polynomials of frequency or its logarithm. In Table \ref{tab:fg_model}
we list some foreground models used in the literature along with short names \citep{2008PhRvD..78j3511P, 2010PhRvD..82b3006P, 2012MNRAS.419.1070H,Bowman2018Naturea,Hills2018Nature}. 

\begin{table}[ht]
	\centering
	\caption{Foreground fitting models, $\nu_c = 75 \MHz$. }
	\label{tab:fg_model}
	\begin{tabular}{ll} 
		\hline
		Model & Fitting Function               \\
		\hline 
		(a) 5-terms {\it Poly} & 
		$\displaystyle{ T_{\rm{F}}(\nu)=\sum_{n=0}^{4} a_{n} \left(\frac{\nu}{\nu_{\rm{c}}}\right)^{n-2.5} }$  \\
	    (b) 7-terms {\it Poly} & 
	   	$\displaystyle{T_{\rm{F}}(\nu)=\sum_{n=0}^{6} a_{n} \left(\frac{\nu}{\nu_{\rm{c}}}\right)^{n-2.5} }$ \\
		(c) 3-terms {\it LogPoly} & $\displaystyle{	T_{F}(\nu) = \left(\frac{\nu}{\nu_{c}}\right)^{-2.5} \sum_{n=0}^{2} a_{n}\left(\log \frac{\nu}{\nu_{c}}\right)^{n} }$ \\
		(d) 5-terms {\it LogPoly} &  $\displaystyle{	T_{F}(\nu) = \left(\frac{\nu}{\nu_{c}}\right)^{-2.5} \sum_{n=0}^{4} a_{n}\left(\log \frac{\nu}{\nu_{c}}\right)^{n} }$   \\
		(e) 3-terms {\it LogLogPoly} &  $\displaystyle{ \log T_{\rm{F}}(\nu)=\sum_{n=0}^{2} a_{n} \left(\log \frac{\nu}{\nu_{\rm{c}}}\right)^n }$\\
		(f) 5-terms  {\it LogLogPoly} &   $\displaystyle {\log T_{\rm{F}}(\nu)=\sum_{n=0}^{4} a_{n} \left(\log \frac{\nu}{\nu_{\rm{c}}}\right)^n }$ \\
		\hline
	\end{tabular}
\end{table}

\begin{figure}[htb!]
 	\centering
    \includegraphics[width=0.9\columnwidth]{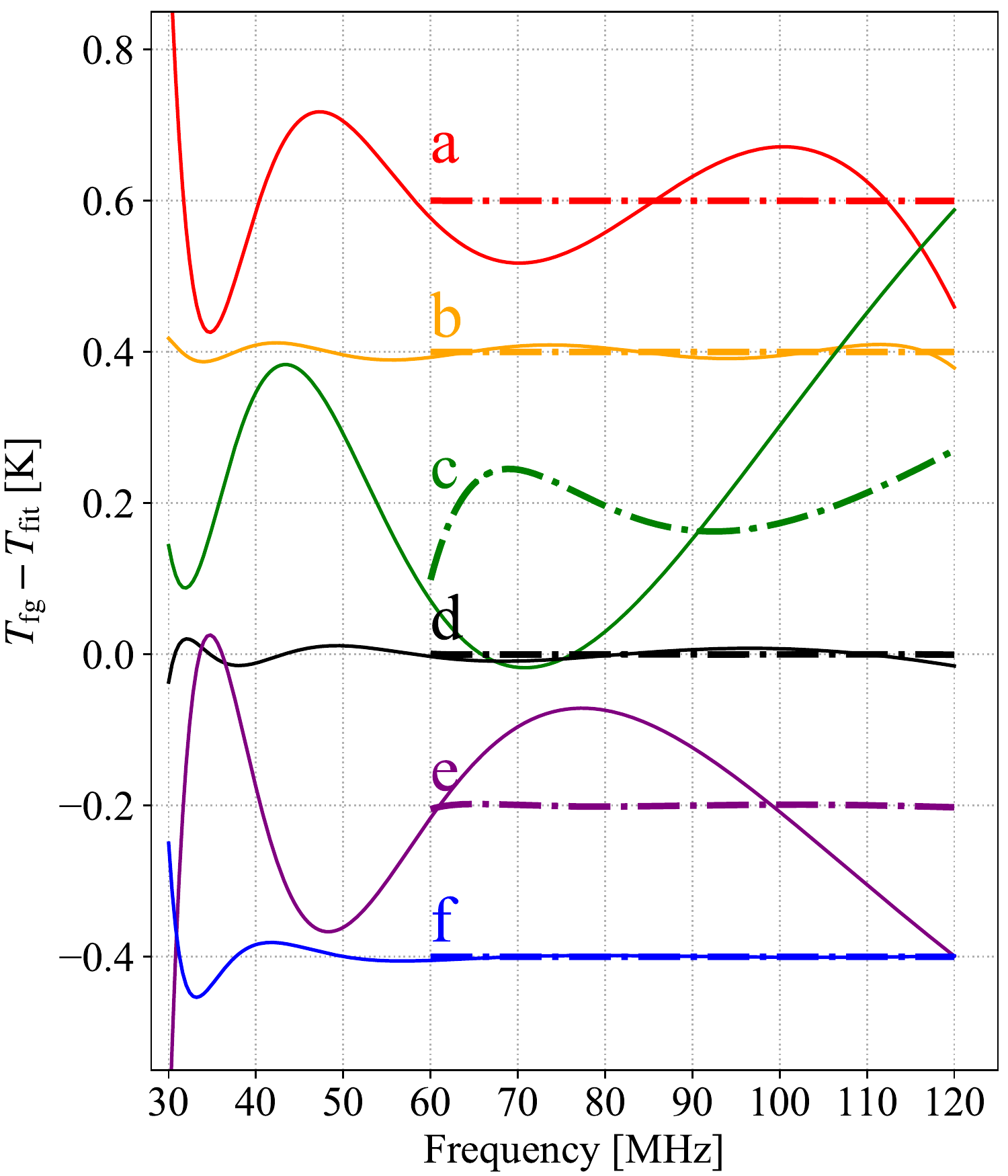}
    \caption{The residuals after fitting the six models listed in Table~\ref{tab:fg_model}, in the range of 30 -- 120 MHz (solid lines) and 60 -- 120 MHz (dash-dotted lines). The lines are shifted vertically for clarity. (a) five-term Poly (+0.6 K), (b) seven-term Poly (+0.4 K), (c) Three-term LogPoly (+0.2 K), (d) Five-term LogPoly, (e) Three-term LogLogPoly (-0.2 K), and (f) five-term LogLogPoly (-0.4 K). 
    }
    \label{fig:foreground_fitting}
\end{figure}

Fig.~\ref{fig:foreground_fitting} shows the residuals of fitting the global spectrum using the models listed in Table~\ref{tab:fg_model}. 
Here we show both a fit in the frequency range of 30 -- 120 MHz (solid lines), and a fit in the frequency range of 60 -- 120 MHz (dash lines) for each model. The RMS of residuals of the (a) - (f) models are 78.37 mK (0.06 mK), 7.58 mK (0.01 mK), 172.74 mK (33.69 mK), 8.14 mK (0.17 mK), 117.60 mK (1.21 mK), 16.08 mK(0.01 mK) for the solid lines (dash-dotted lines) respectively.

As shown by the figure and the residues, the primary source of foreground, the galactic synchrotron radiation, has a nearly power law form spectrum, which can be easily fitted with few terms in the polynomial expansion. However, the ISM free-free absorption at the low frequency breaks the power law. So when the fitting is expanded to the 30 -- 120 MHz frequency range, the five-term {\it Poly} and three-term {\it LogLogPoly} models which worked well in the 60--120 MHz range become inadequate. 
The five-term {\it LogPoly}, five-term {\it LogLogPoly} and seven-term {\it Poly} models can fit well in all cases. We take the five-term {\it LogPoly} model as our fiducial model for the foreground below.

\subsection{Models of the 21 cm signal}
\label{subsec:21 cm}

We consider here two models of the 21 cm global signal: one is form adopted by the EDGES experiment in \cite{Bowman2018Naturea}, another is a Gaussian fit for the 21 cm through with adjustable amplitude and width \citep{Hills2018Nature}. 

{\bf EDGES 21 cm model.}  This is motivated by the EDGES experiment observation. We assume the signal is 
\begin{equation}
T_{21}(\nu)=-A\left(\frac{1-\rm{e}^{-\tau \rm{e}^{B}}}{1-\rm{e}^{-\tau}}\right)
\label{eq:T21_flat}
\end{equation}
where
\begin{equation}
B=\frac{4\left(\nu-\nu_{0}\right)^{2}}{w^{2}} \log \left[-\frac{1}{\tau} \log \left(\frac{1+\rm{e}^{-\tau}}{2}\right)\right]
\end{equation}

{\bf Gaussian model.} As demonstrated in  \citet{Cohen2018MonNotRAstronSoc},  different astrophysical parameters can produce a variety of global 21 cm spectrum with absorption trough depths in the range of $-200\; \mr{mK} < T_b < -25\; \mr{mK}$. \cite{Xu2021ArXivE-Prints} also took into account the inevitable non-linear density fluctuations in the IGM, shock heating and Compton heating, and found that these effects can reduce the maximum absorption signal by 15\% at redshift 17.  Here we use a simpler three-parameter Gaussian form as given in Eq.~(\ref{eq:gaussian}) as our global signal model.
\begin{equation}
	T_{21}(\nu)=-A e^{\frac{-(v-\nu_0)^{2}}{2 w^{2}}}
	\label{eq:gaussian}
\end{equation}

\begin{figure}[tbp]
	\centering
    \includegraphics[width=\columnwidth]{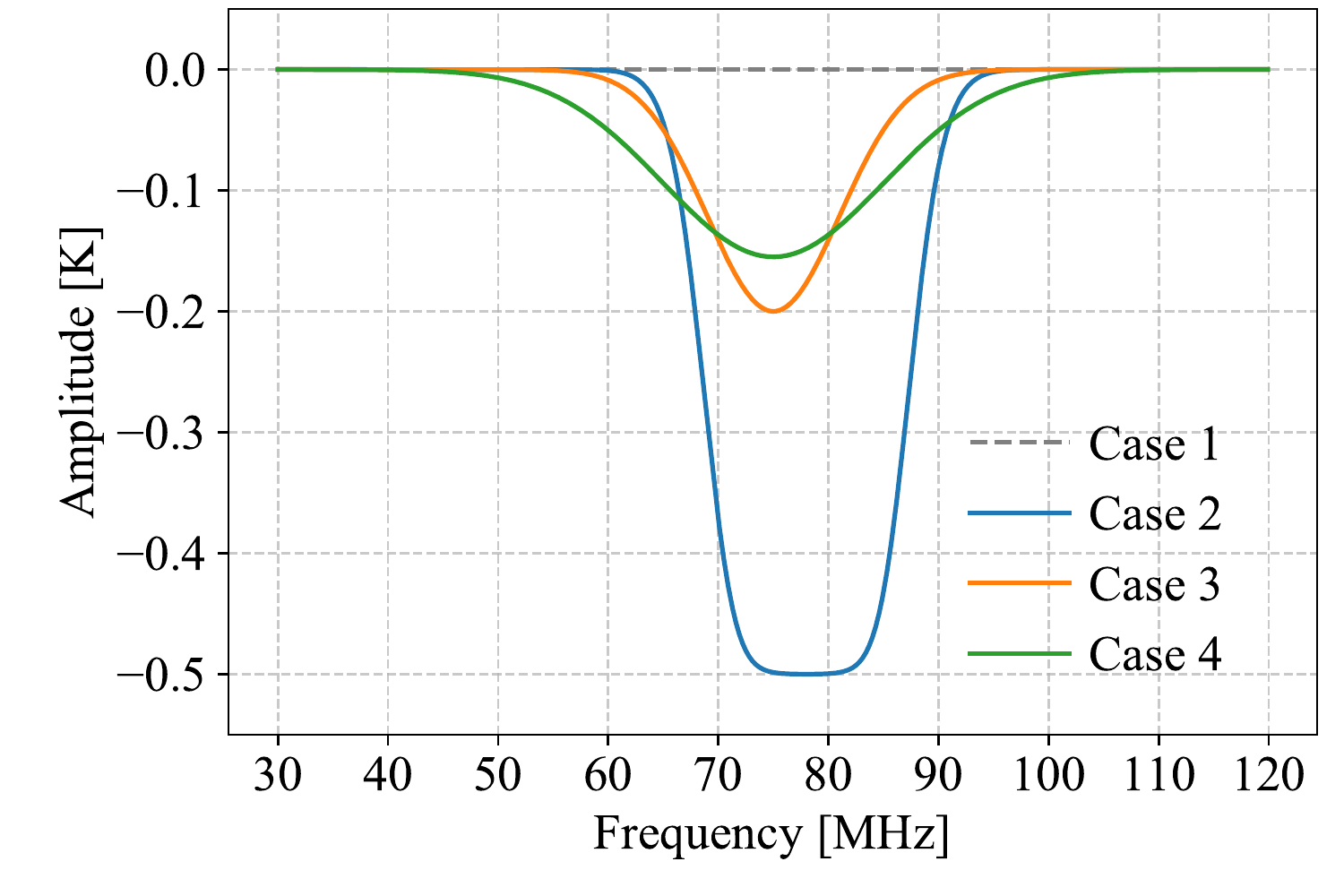}
    \caption{The 21 cm signal for different scenarios at 30 -- 120 \MHz. Case 1  (no 21 cm signal) is shown by the grey dashed line, Case 2 (EDGES signal) is shown by the blue solid line,  Case 3 and Case 4 (two Gaussian forms) are shown by the yellow and green solid lines respectively.}
    \label{fig:21 cm}
\end{figure}

Then we consider four cases for the global 21 cm signal profile:

{\bf Case 1} - No detectable 21 cm global signal.

{\bf Case 2} - The EDGES 21 cm signal as given by Eq.(\ref{eq:T21_flat}), with $A = 0.5$ K, $\nu_0 = 78.0 \MHz$, $w = 19.0 \MHz$, $\tau = 7$.

{\bf Case 3} - A Gaussian signal as given by Eq.(\ref{eq:gaussian}), with $A = 0.20 \K$, $\nu_0 = 75.0 \MHz$, $w=6.0 \MHz$.

{\bf Case 4} - A Gaussian signal as given by  Eq.(\ref{eq:gaussian}), with $A = 0.155 \K$, $\nu_0 = 75.0 \MHz$, $w=10.0 \MHz$.

Fig.~\ref{fig:21 cm} illustrates the 21 cm global signal for these four scenarios. 

\subsection{Thermal Noise}
\label{subsec:noise}
\begin{figure}[h]
    \centering
    \includegraphics[width=0.75\columnwidth]{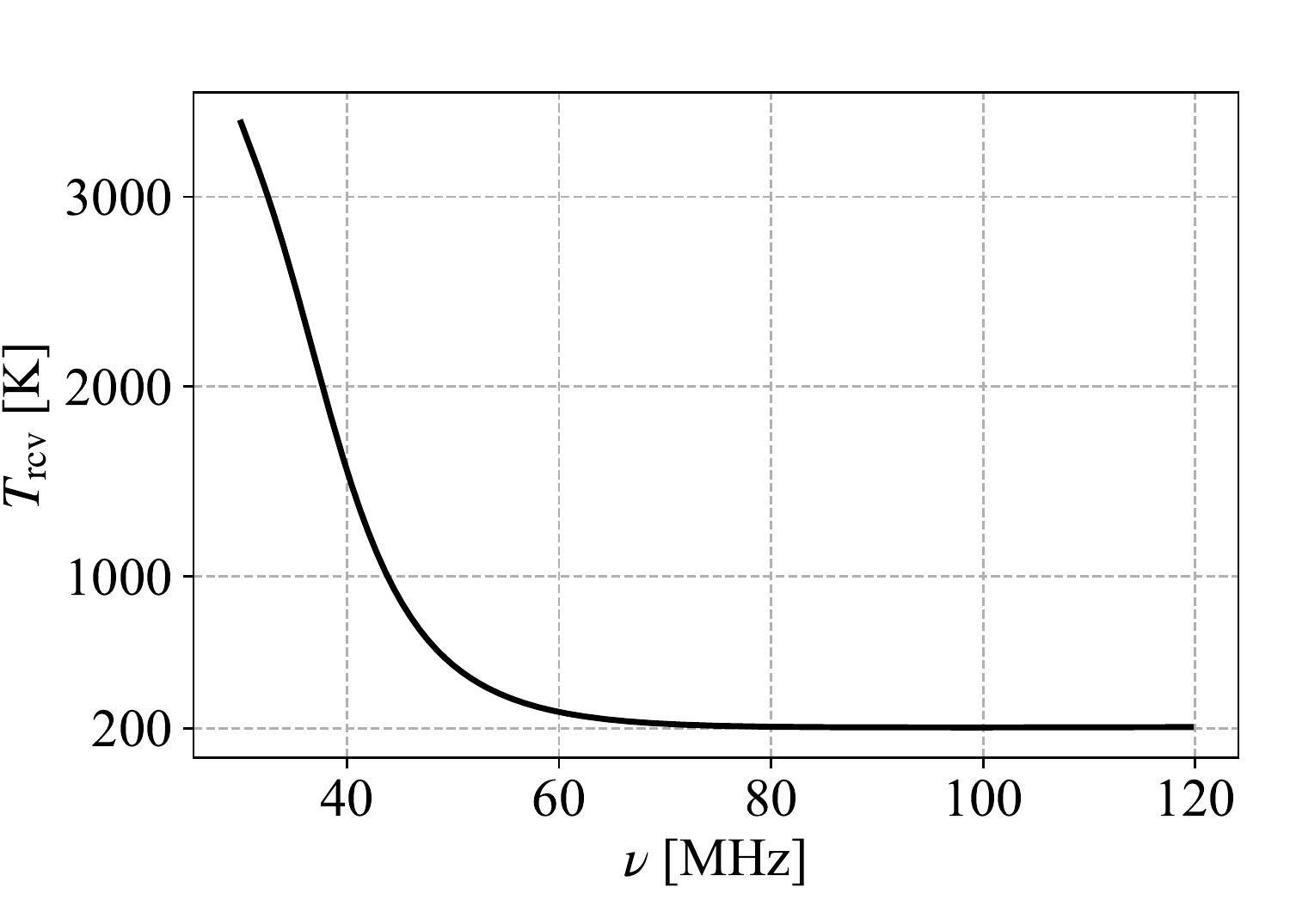}
    \caption{The equivalent receiver temperature.}
    \label{fig:T_rcv}
\end{figure}

The thermal noise on the observed signal is \begin{equation}
	\sigma_n=\frac{T_{\rm{sys}} }{\sqrt{N \Delta \nu t_{\rm{int}}}},
	\label{eq:noise}
\end{equation}
where $N$ is the number of independent measurements, $\Delta \nu$ is the channel bandwidth which we set as 0.4 MHz, and $t_{\text{int}}$ is the integration time. 
The system temperature is given by
\begin{equation}
T_{\rm{sys}} = T_{\rm{sky}} + T_{\rm{rcv}},
\end{equation}
where the average sky temperature as is given by 
\begin{equation}
 T_{\mr{sky}} \sim  2300 \left( \frac{\nu}{75 \text{MHz}} \right)^{-2.5}. 
 \label{eq:T_sky}
 \end{equation}
Note that while it is important to account for the ISM absorption when fitting the low frequency spectrum, which would affect the extracted 21 cm signal several orders of magnitude lower than the foregrounds, the absorption effect makes little difference when estimating the thermal noise. Therefore, a simple power-law model is assumed here.
$T_{\rm{rcv}}$ gives the equivalent antenna temperature induced by the noise in the receiver. As the antenna and receiver impedance are mismatched, the actual power goes in to the receiver is $(1-|\Gamma|^2)P_{\rm ant} $, where $\Gamma$ is the reflection coefficient between the antenna and receiver, and $P_{\rm ant}$ is the power of the antenna output. So 
\begin{equation}
T_{\rm{rcv}}= \frac{T_{\rm{rcv}}^0}{1-|\Gamma|^2},
\end{equation}
where $T_{\rm{rcv}}^0$ is the noise temperature of the receiver.
For our current receiver design, we have measured $T_{\rm{rcv}}^0 \approx 200 \K$ in the given band. Based on antenna simulation, we find that
\begin{equation}
\frac{1}{1-|\Gamma|^2} \approx 16.809 \exp \left(-\frac{\left(\frac{\nu}{\nu_{\rm{c}}}-0.4\right)^{1.277}}{0.0808}\right)+1.034.
\end{equation}
We plot $T_{\rm{rcv}}$ in Fig.\ref{fig:T_rcv}. It is about 200 K near 80 MHz.  

With such a system temperature, for a single antenna ($N=1$),  after about 24 hours (10 orbits) of integration time, the RMS noise level is $\leq 15$ mK at 75 \MHz, assuming a channel bandwidth of 0.4 \MHz. Even if we assume that only the data obtained during the ``doubly good time'', when both the Earth and Sun are shielded, can be used, which is about 10\% of the total orbit time, it takes only 10 days for the thermal noise to be suppressed down to this level according to Eq.(\ref{eq:noise}). 

The detailed study of the global spectrum in the 1 -- 30 MHz band will be presented elsewhere, but here we note in passing that at 25 MHz, the sky temperature is about $3.6\times 10^4 \K$, according to Eq.~(\ref{eq:T_sky}). After a full three year of integration, the thermal noise of a single antenna could be suppressed to a level of about 10 mK for 0.4 MHz channel bandwidth. There are 5 -- 8 satellites in the DSL mission, and each is equipped with 3 pairs of antennas of different polarization. So the DSL mission has the potential to measure the 21 cm global spectrum from the dark ages and cosmic dawn. However, we note that this estimate accounts only for the thermal noise, a number of other factors or systematic errors, such as the self-generated RFI, stability, dynamic range, and spectral response of the system will probably limit the precision of the measurement. It is necessary to design the system such that these systematic errors to be smaller than the expected 21 cm signal.

\subsection{Beam}
\label{subsec:beam}

The global spectrum in the 30 -- 120 MHz band will be measured by an antenna on a single satellite in the DSL mission. The primary requirement of the design is to have low chromaticity, i.e. its beam pattern should be as frequency-independent as possible, to avoid convolving the spatial structure of the sky into the frequency domain. 
This can be achieved by using an antenna which is electrically small, i.e. with a physical size smaller than the half-wavelength, so that the resonance frequency is above the observational band. The small size of the antenna is also consistent with the requirement on the payload of the mission. 

\begin{figure}[!thbp]
	\centering
    \includegraphics[width=0.8\columnwidth]{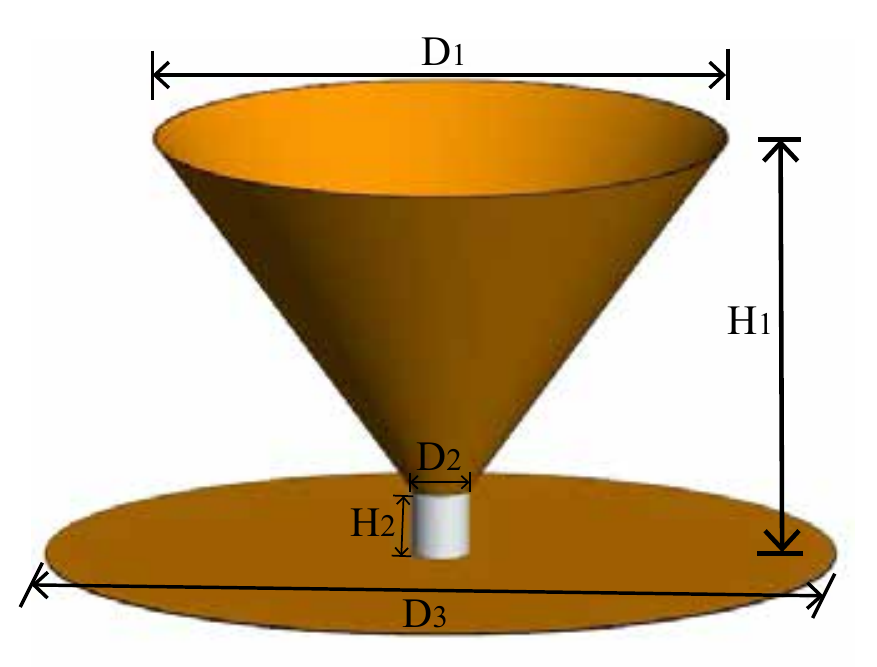}
    \caption{The disc-cone antenna. The size parameters are $D_1 = 40 \cm$, $D_2=4 \cm$, $H_1 = 30.5 \cm$, and $H_2 = 4.5 \cm$. The diameter of the optimal metallic reflector $D_3$ is 55 cm in our fiducial model.}
    \label{fig:antenna}
\end{figure}  

\begin{figure}[tbp]
\centering
\includegraphics[width=0.8\columnwidth]{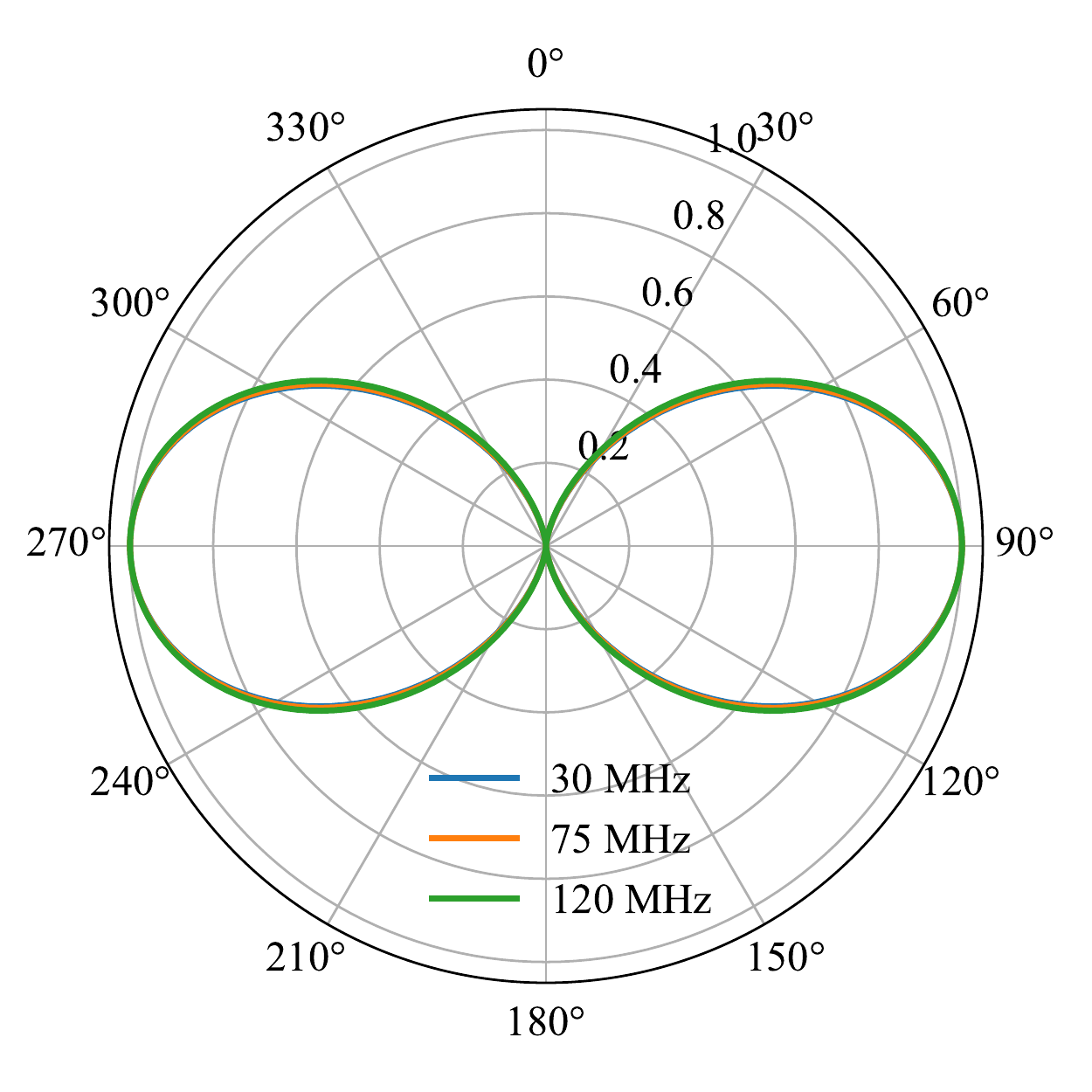}
\includegraphics[width=0.8\columnwidth]{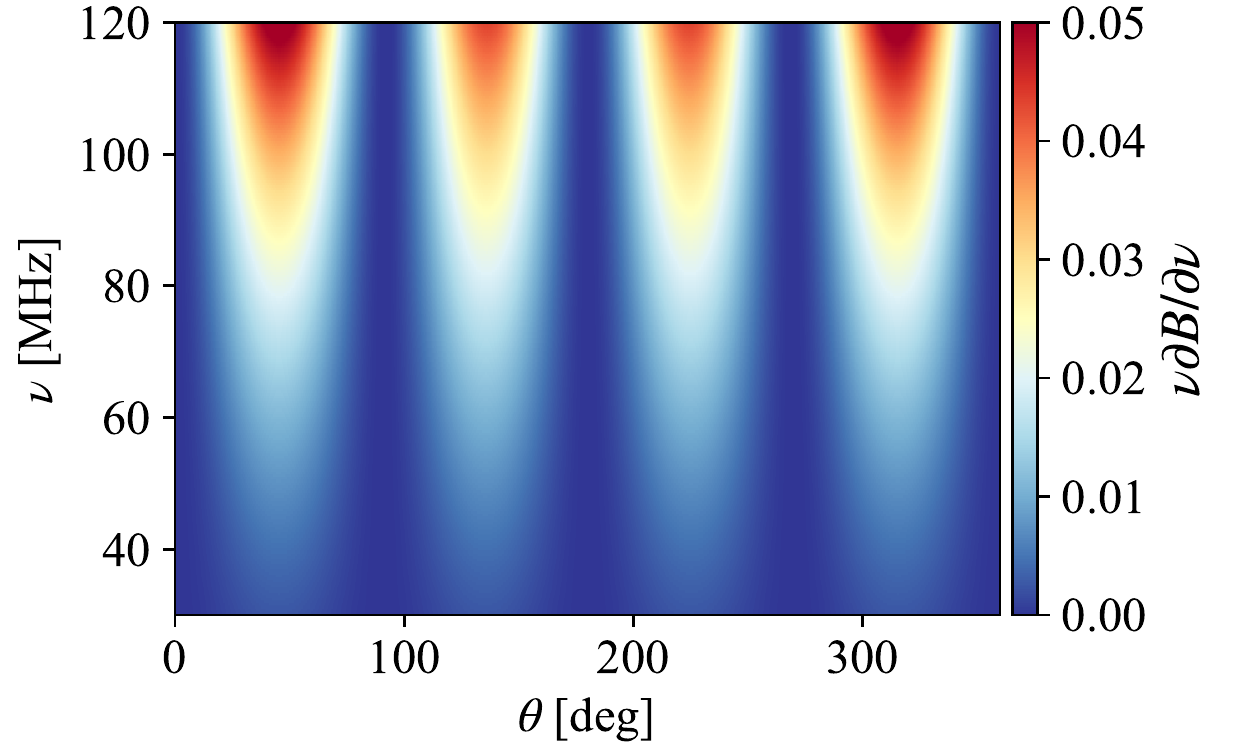}
\caption{The cross section of the beam profiles at a constant $\phi$ for the fiducial model at three frequencies (top) and the corresponding frequency gradients 
at various directions and frequencies (bottom).}
\label{fig:beam_fid}
\end{figure}

In the present design, we consider a disc-cone antenna model, which is composed of a radiating cone and a reflective disk, as shown in Fig.~\ref{fig:antenna}. The size parameters of the cone and disc are labelled on the figure. The design of this antenna will be presented elsewhere, here we take a particular design as our fiducial model. The beam of this antenna is simulated using the software FEKO\footnote{\url{Altair FEKO - https://altairhyperworks.com/product/FEKO}}. The cross section of the axial-symmetric beam profiles, cut at an arbitrary $\phi$ with varying $\theta$, is shown in the top panel of Fig.~\ref{fig:beam_fid}, the actual beam in the 3D space can be generated by rotating the cross section figure around the central axis ($\theta=0\degree$).
The beam pattern of this electrically small disc-cone antennas is quite similar to that of a short dipole; there is a single lobe of the beam (it peaks at $\theta = 90^\circ$ and $270^\circ$ on the cross-section plot). To see the effect of frequency variation of the beam more clearly, we also show the logarithmic frequency gradient of the beam $\partial B(\nu,\theta)/\partial \log\nu= \nu\partial B(\nu,\theta)/\partial\nu$ in the bottom panel of Fig.~\ref{fig:beam_fid}. The gradient for this disc antenna is fairly small. In the next section, we shall mainly use this antenna beam model, but to explore how the chromatic beam affects the global spectrum measurement result, we shall also consider beam models which are more frequency-dependent.

\section{Simulation and Results}
\label{sec:simulation}

With the set up described above, we simulate the observed global spectrum. First we investigate how the measured global spectrum 
varies as a function of time, as the satellite moves to different parts of the orbit. Next we estimate the measurement error after some time of observation. 
Then we consider how the systematic errors,
such as the chromatic beam effect, and the non-flat spectral response of the system would affect the 21 cm signal measurement.

\subsection{Variation of Observing Position}
\label{sec:sim_pos}
\begin{figure}[tbp]
    \centering
    \subfigure[A]{
    \includegraphics[width=0.45\columnwidth]{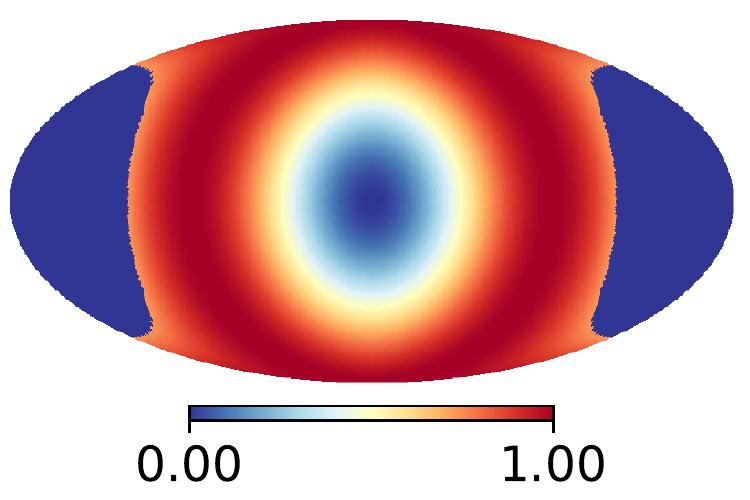}}
    \subfigure[B]{
    \includegraphics[width=0.45\columnwidth]{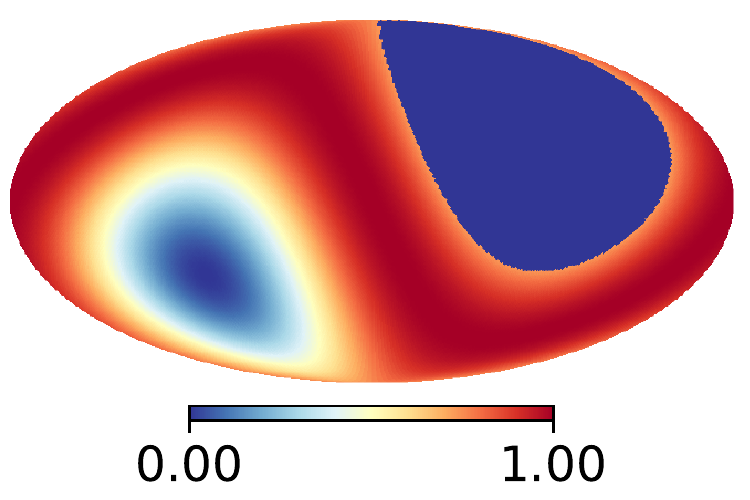}}\\
    \subfigure[C]{
    \includegraphics[width=0.45\columnwidth]{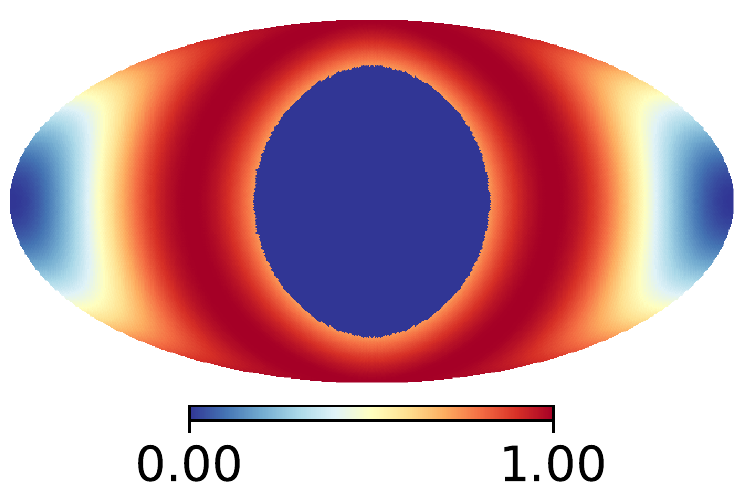}}
    \subfigure[D]{
    \includegraphics[width=0.45\columnwidth]{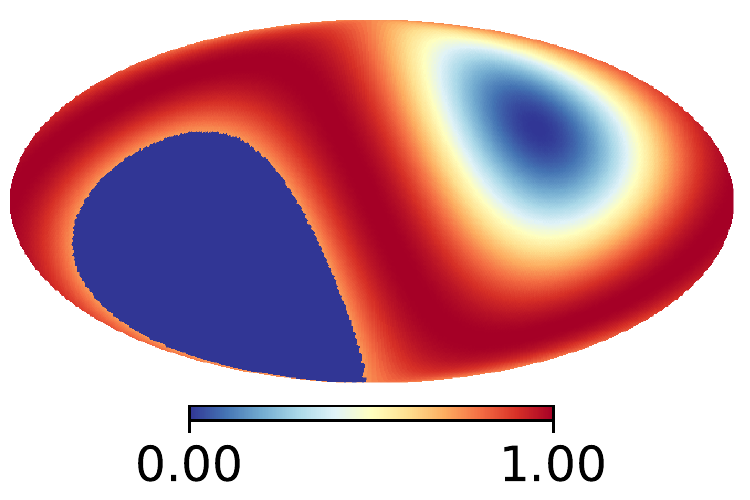}}
    \centering
    \caption{The beam seen from the four positions A, B, C and D, which are located $90^\circ$ apart, around one orbit. The value is normalized to fall in  the range [0, 1] .}
    \label{fig:signal_abcd}
\end{figure}

As there are large scale variations in the foreground, and the Moon blocks different parts of the sky for the satellite orbiting to the different 
positions. We first check how much the variation is for different locations. For this purpose, we select four points, marked as A, B, C, and D, 
which are located  $90^\circ$ apart  on the lunar orbit. These are essentially the cross-over points of the orbit with the ecliptic plane, and the 
highest and lowest points away from the elliptic plane along the orbit.  
The observed global spectrum at these points are the averaged sky temperature as weighted by  $S(\mathbf{n}) B(\mathbf{n})$ at these locations. 
In Fig.~\ref{fig:signal_abcd}, we show the 
product of the antenna beam and Moon blockage function, $S(\mathbf{n}) B(\mathbf{n})$, at the  four locations. 
As expected from the symmetry, the pair A and C have exactly opposite patterns, and so does the pair B and D.

\begin{figure}[htbp]
	\centering
    \includegraphics[width=\columnwidth]{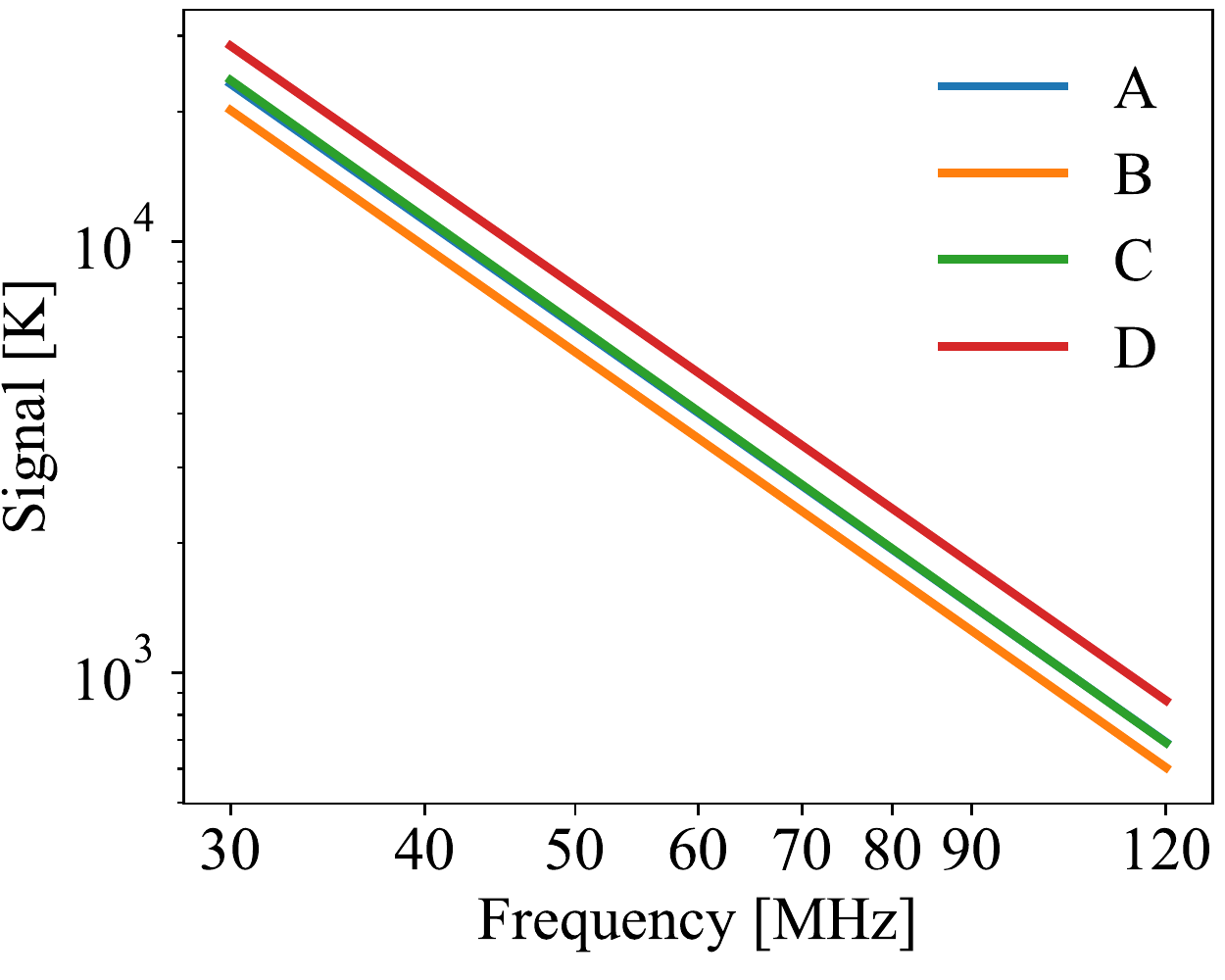}
    \caption{Observed foreground spectra at the positions A, B, C, and D.  }
    \label{fig:satellite_abcd}
\end{figure}

We then calculate the foreground spectra as would be observed at these points with our fiducial disc-cone antenna model described above. In Fig.~\ref{fig:satellite_abcd}, we plot the global spectra in the 30 -- 120 MHz range. There are obvious differences in the amplitude for the four locations, but the shapes of the spectra are still quite similar with each other, and the spectra from locations A and C almost coincide. 

\begin{figure}[htbp]
    \centering
    \includegraphics[width=.99\columnwidth]{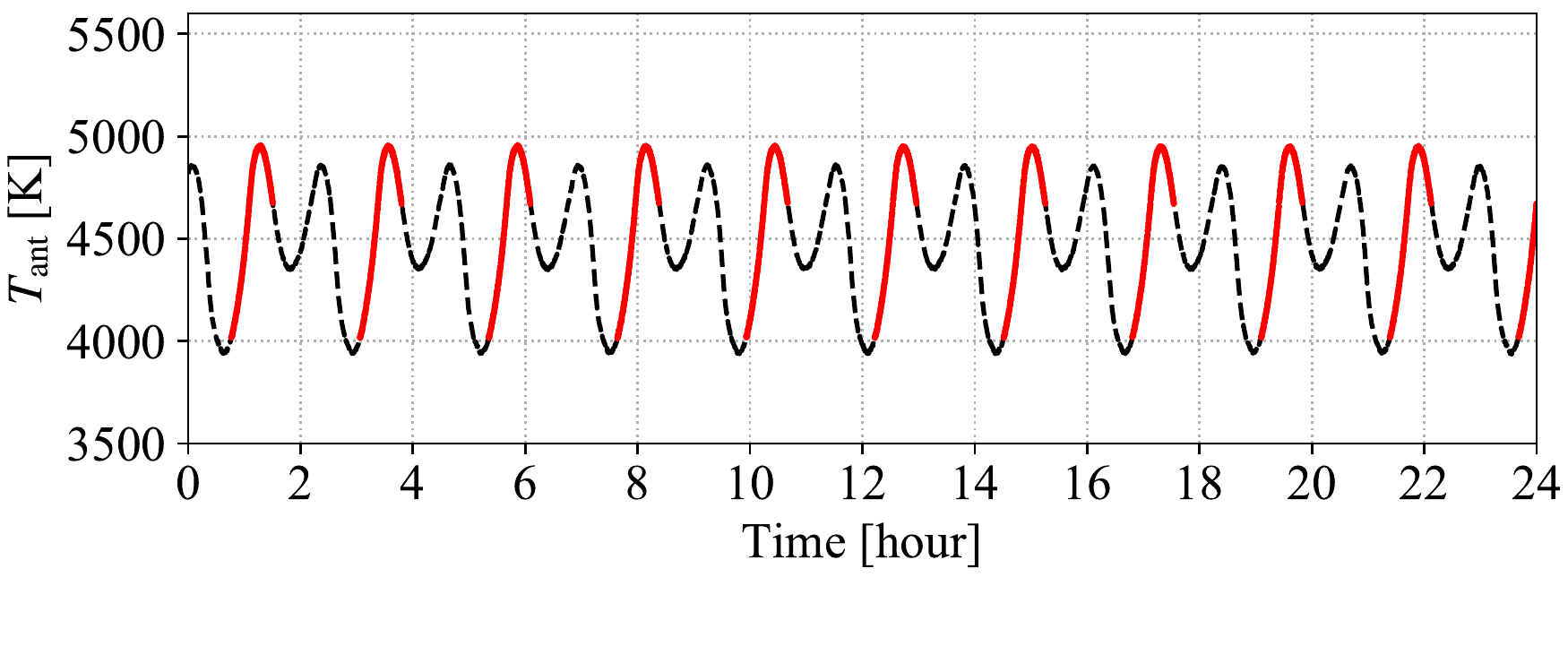}
    \caption{Signal received at 60 MHz over 24 hours observation. The black dashed line is the spectra for 1 day continuous observation, and the red solid line accounts for the 1/3 effective observation time, during which the satellite is shielded by the Moon from the Earth.} 
    \label{fig:signal_A}
\end{figure}

Just as shown by the spectra at these four points, there is large variation in the total intensity.
In Fig.~\ref{fig:signal_A}, we plot the simulated intensity variation at 60 MHz, as the satellite taking an observation of 24 hours. In this plot, the red solid line mark the part of data obtained when the Earth is shielded, while the black dashed line are for the rest of time. The figure shows that the signal received at a fixed frequency has a regular periodic variation, with a period of 8246 seconds corresponding to one cycle of the orbit, and an amplitude of $\sim 10\% $ level.

\begin{figure}[htbp]
    \centering
    \includegraphics[width=0.95\columnwidth]{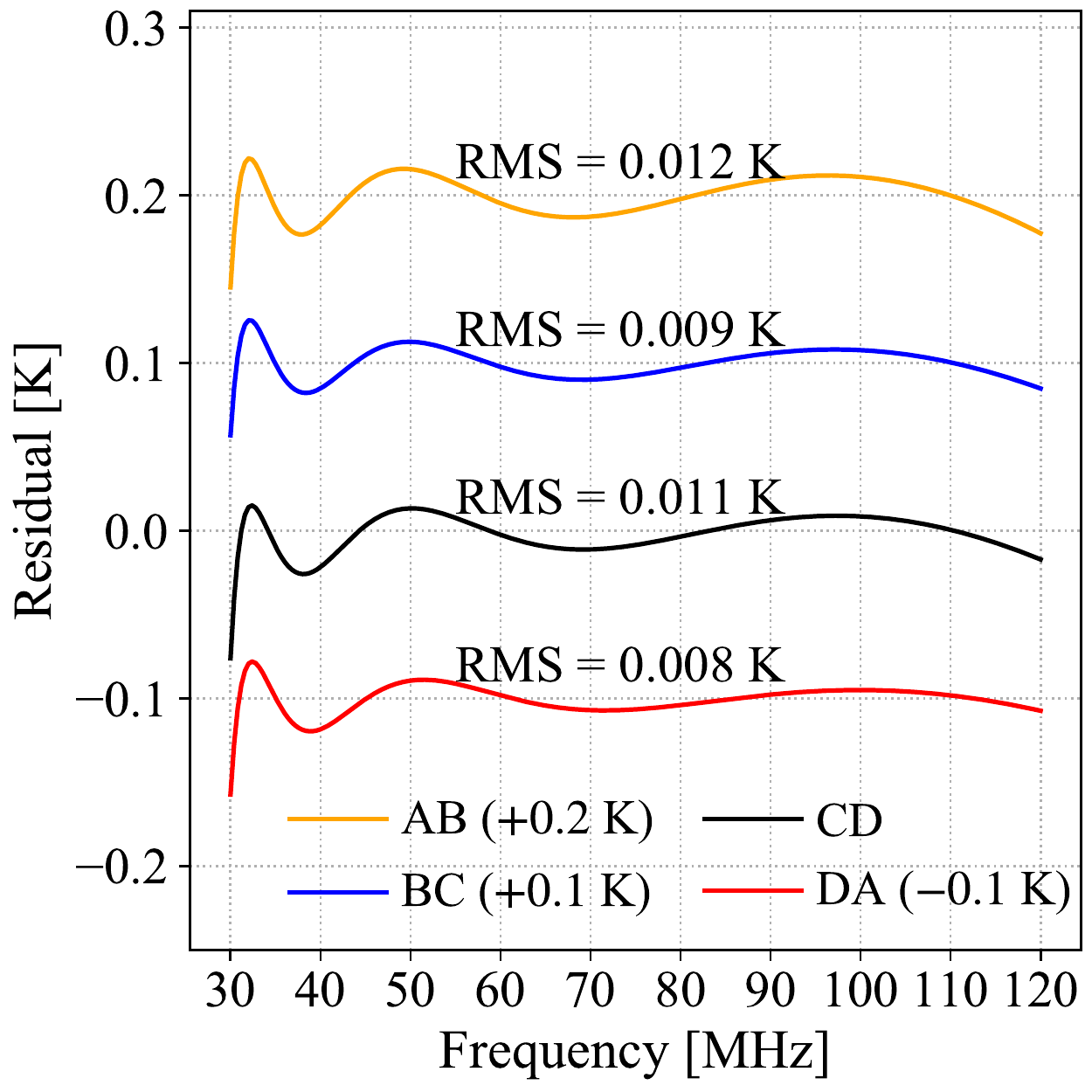}
    \caption{Residuals of the observed foreground spectra for the four segments of orbit, after fitting the five-terms {\it LogPoly} model. 
    }
    \label{fig:resi_4p}
\end{figure}

Actually the global spectrum is measured over a period of time as the satellite orbits the Moon. We can compute the sky average for a segment of the orbit.
Despite large variations in the overall amplitude of the antenna temperature at different parts of the orbit, as shown in Fig.~\ref{fig:signal_A}, the variation in the shape of the spectrum is not large.

\begin{figure}[thbp]
	\centering
    \includegraphics[width=0.9\columnwidth]{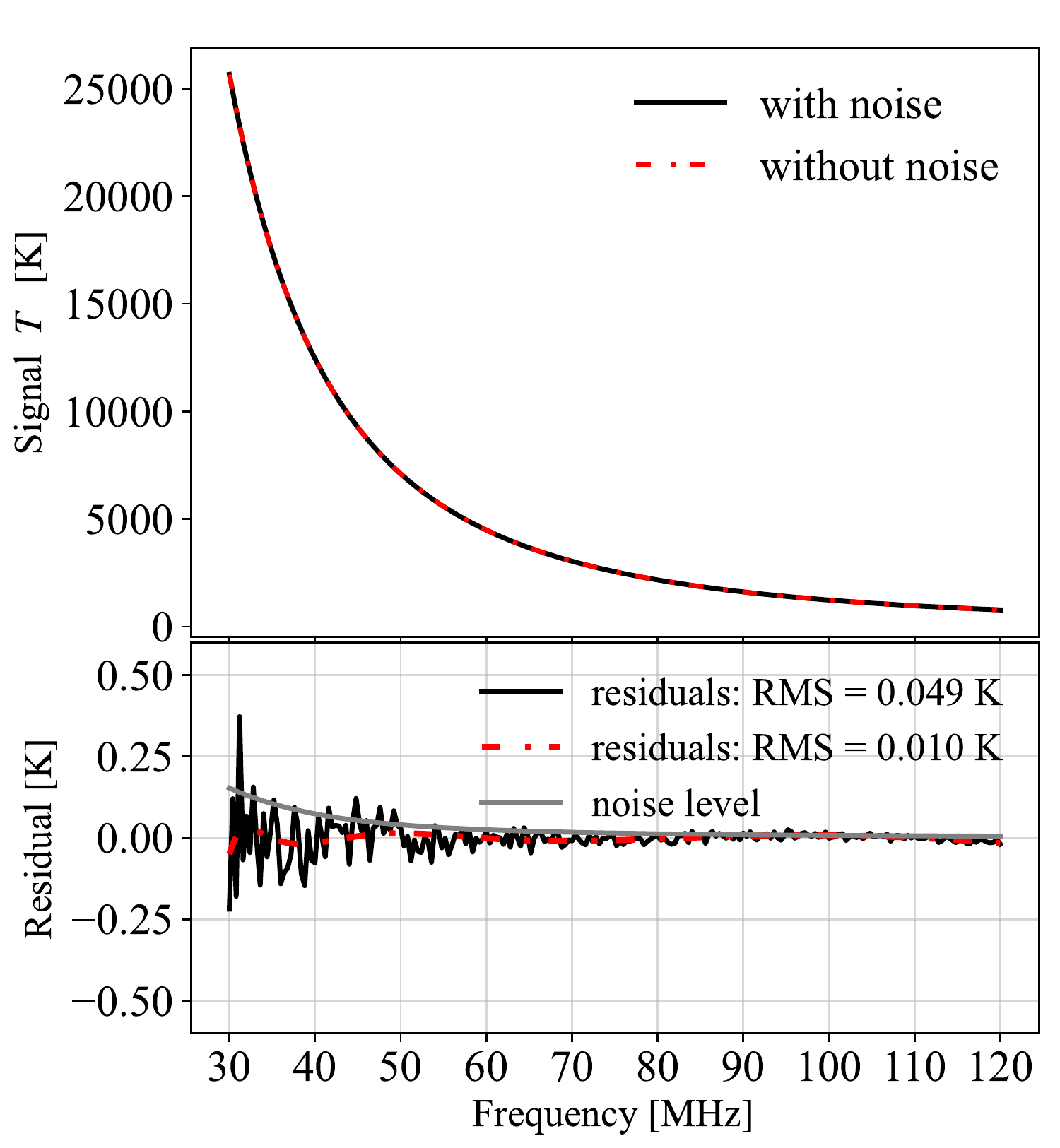}
    \caption{Top: The average spectrum for 10 orbits of effective observation (total 30 orbits but 1/3 shielded by the Moon from the Earth). Bottom: The residuals of foreground fitting for the noise-included (black solid line) and noise-free (red dash-dotted line) cases. } 
    \label{fig:s1_result}
\end{figure}

We fit each of the integrated spectrum for the AB, BC, CD and DA segments of the orbit separately with the five-terms {\it LogPoly} foreground model, and plot the residuals in Fig.~\ref{fig:resi_4p}. Here the input sky model includes only the foreground, while the mock observation is performed with simulated noise, Moon blockage, and the fiducial model of antenna response.  
For clarity, the results are vertically shifted by different amounts in the figure. We find that the beam-weighted signals can be reconstructed with a high accuracy. The four segments yield similar residuals, and these residuals (0.012 K, 0.009 K, 0.011 K, 0.008 K, respectively) are all noticeably smaller in magnitude than the expected global 21 cm signal in typical cosmic dawn models, showing that despite the variation of foreground with sky direction, this would not seriously affect the 21 cm signal extraction.

\subsection{Projected Error}
In Fig.~\ref{fig:s1_result}, we plot the averaged spectrum at 30 -- 120 MHz after a total of 30 orbits, but take observation only when the Earth is shielded which is about 1/3 effective time. In the same figure we also plot the fitting residual of the foreground model, for both the noise-free case and with noise given at a level corresponding to $t_\text{obs}= 10$ orbits. The data is assumed to be taken with our fiducial antenna model.

\begin{figure}[htbp]
	\centering
    \includegraphics[width=1.0\columnwidth]{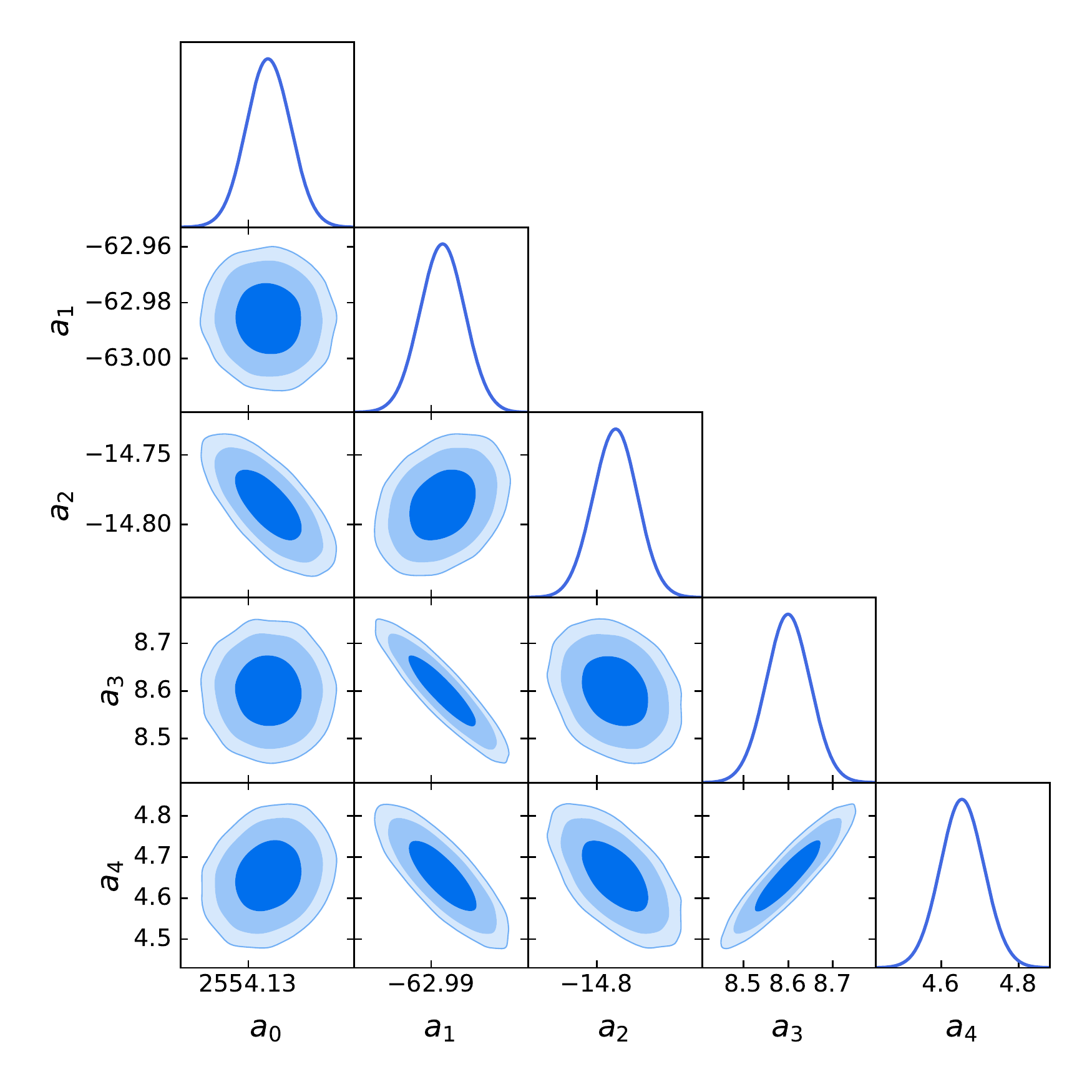}
    \caption{Foreground model parameter fitting results for 10 orbits observation. We plot the contour maps with 1$\sigma$, 2 $\sigma$ and 3 $\sigma$ for {\it LogPoly} foreground coefficients $a_0$,  $a_1$,  $a_2$,  $a_3$,  $a_4$. The solid lines are the 1-D PDFs of the parameters.  }
    \label{fig:s1_mcmc}
\end{figure}

We find that the {\it LogPoly} model is able to account for the global spectral structure by expanding around the expected power law spectral index of -2.5. Here $a_0$ is an overall foreground scale factor, $\frac{a_n}{a_0} << 1 \ (n\geq 1)$ is the correction to the typical spectral index of the foreground by capturing anisotropies of the spectral index  and other higher-order spectral corrections, such as the ISM absorption. The five-term {\it LogPoly} model can provide a good fit in this foreground-only sky model, with the foreground model parameters taking reasonable values as shown in Fig.\ref{fig:s1_mcmc}. There is also no serious degeneracy in these parameters. Note that with a micro-satellite on the lunar orbit, our fitting implicitly assumes that the parameter dependencies encode only the foreground properties, and does not account for ionospheric effects.

\begin{figure*}[!htbp]
	\centering
   \subfigure[Results for the EDGES 21 cm model.]{
    \includegraphics[width=0.67\columnwidth]{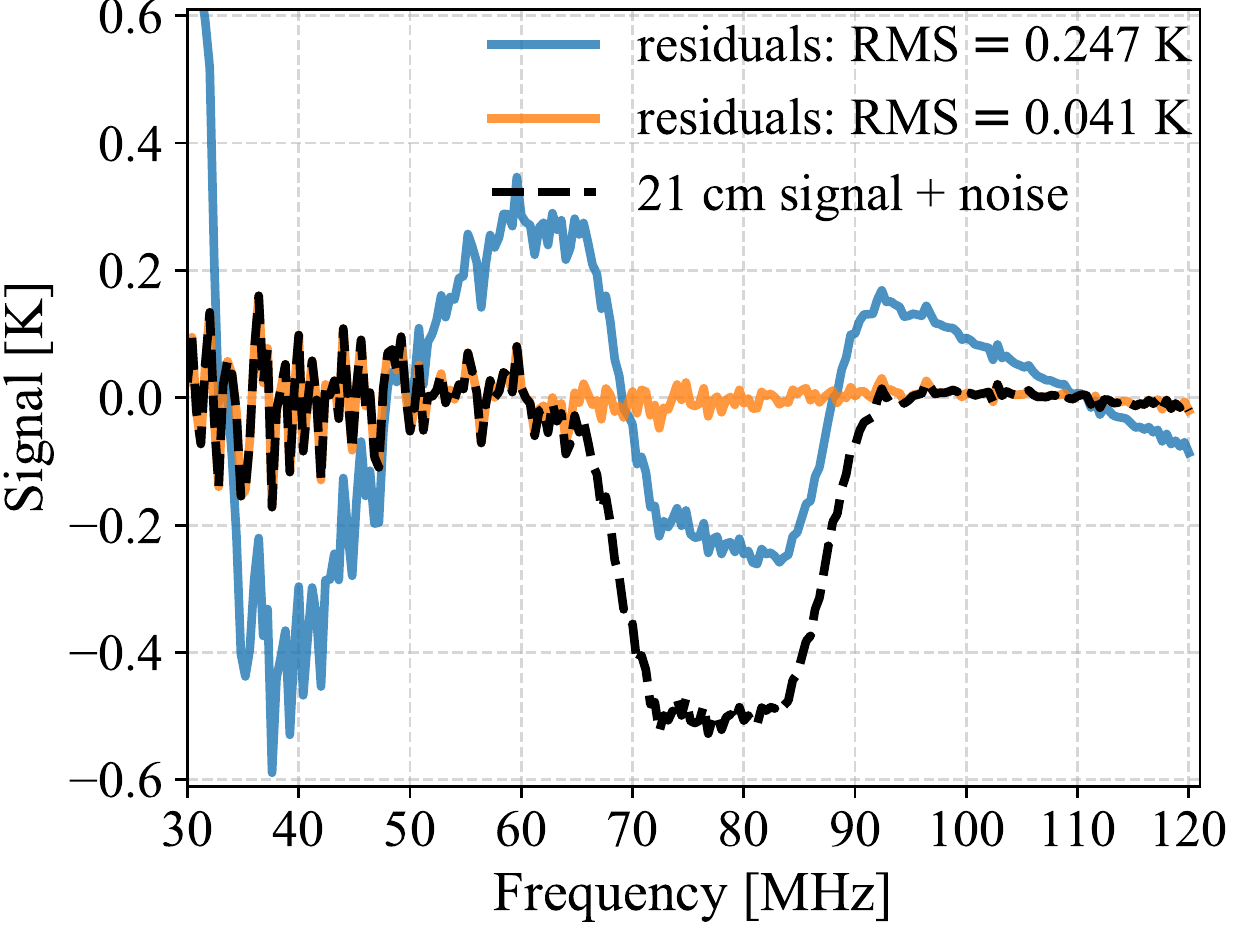}}
    \subfigure[Results for Gaussian 21 cm model ($A=0.2$ \K).]{
    \includegraphics[width=0.67\columnwidth]{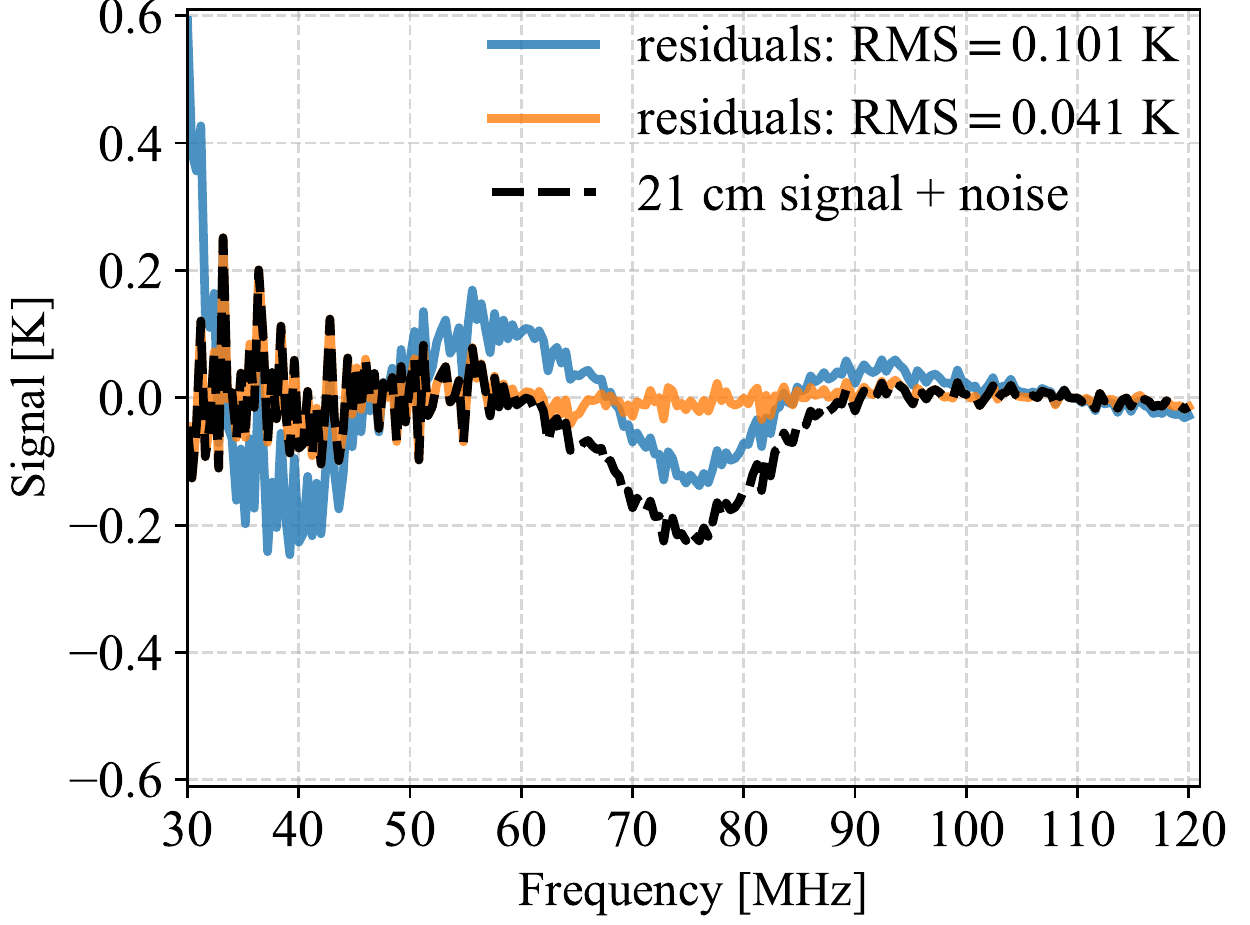}}
    \subfigure[Results for Gaussian 21 cm model ($A=0.155$ \K).]{
    \includegraphics[width=0.67\columnwidth]{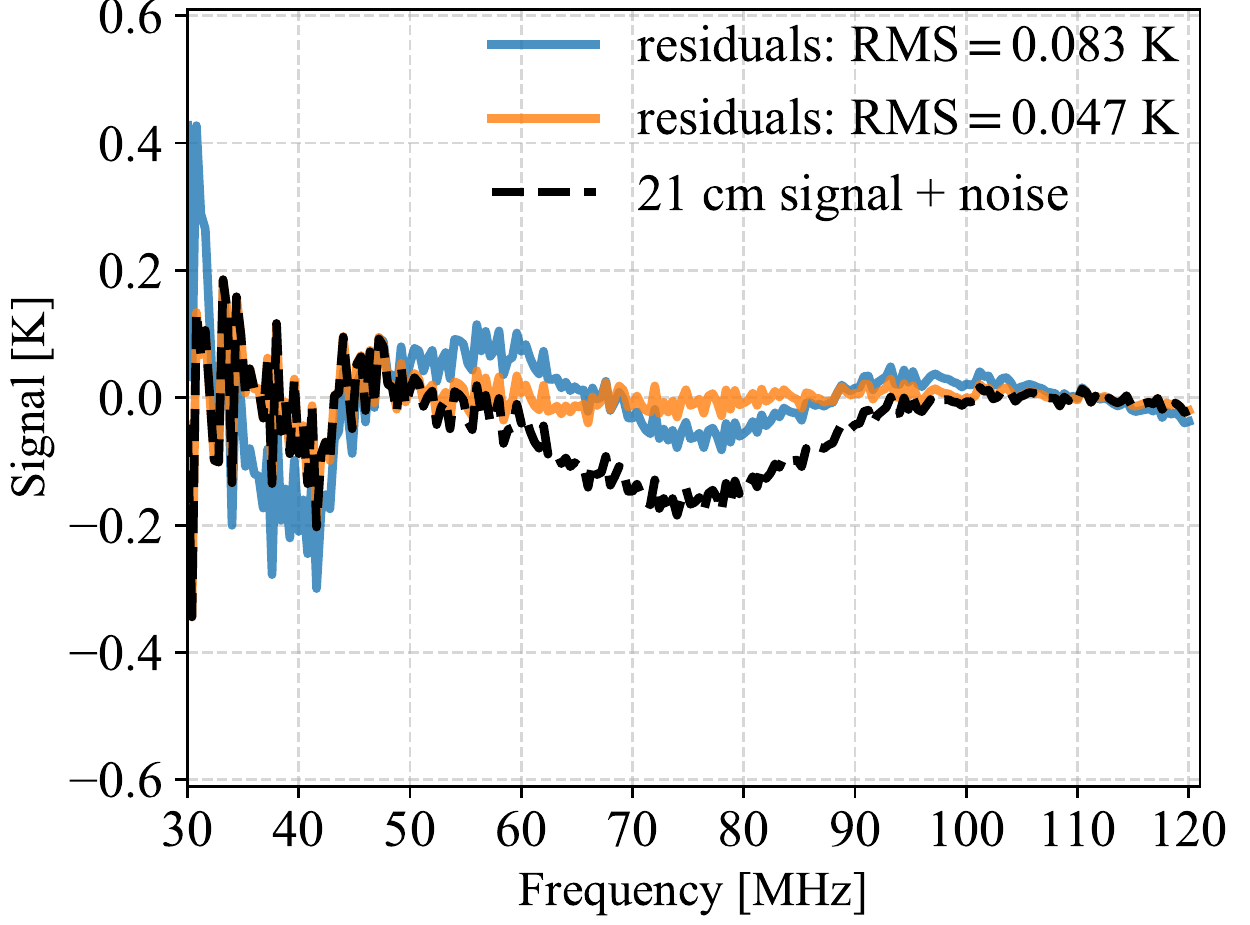}}
    \caption{Fitting results for three 21 cm models after effective 10 orbits observation. The thermal noise is $\sigma_T \sim 0.05$ K. The blue solid lines denote residuals after fitting and removing only the foreground, while the orange lines denote residuals after fitting and removing the combined foreground and 21 cm signal. The black dashed lines denote the recovered 21 cm signals with noise.}
    \label{fig:result}
\end{figure*}

\begin{table*}
\centering
\caption{The best-fit parameters of 21 cm global signal and RMS values for EDEGS model. }
\label{tab:computation_21}
\begin{tabular}{ccccccc}
\toprule[1.5pt]
\makecell[c]{21 cm model} &  \makecell[c]{ $t_{\rm{obs}}$\\ $\text{[orbit]}^1$ } & RMS [K] &  $A$ & $\tau$ & $\nu_0$ & $\omega$ \\
\midrule[1pt]
Input & --  & --& 0.50 & 7.00 & 78.00 & 19.00  \\
\hline
Recovered & 1 & 0.133 & $0.501_{-0.013}^{0.013}$ & $7.2_{-1.1}^{1.5} $ & $77.96_{-0.14}^{0.13} $ & $19.00_{-0.31}^{0.31}$ \\
& 2 & 0.077 & $0.5017_{-0.0093}^{0.0092}$ & $6.84_{-0.82}^{0.94}$ & $ 77.89_{-0.10}^{0.10} $ & $19.00_{-0.23}^{0.22}$\\
& 5 & 0.061 & $0.5001_{-0.0067}^{0.0061}$ & $6.92_{-0.54}^{0.67}  $ & $77.960_{-0.065}^{0.065}$ & $19.02_{-0.15}^{0.14} $\\
& 10 & 0.041 & $0.5004_{-0.0044}^{0.0043}$ & $7.02_{-0.40}^{0.46}$ & $77.979_{-0.045}^{0.046}$ & $ 19.02_{-0.10}^{0.10} $\\
\bottomrule[1.5pt]

\end{tabular}
	\begin{tablenotes}
	\footnotesize
	\item [$^1$] Here as a simple model for the circular orbit, the satellite will orbit the Moon with a period of 8248.7 seconds ($\sim$ 2.3 hr).
	\end{tablenotes}
 \end{table*}

 \begin{table*}
    \centering
    \caption{The best-fit parameters of 21 cm global signal and the RMS values for two Gaussian models.  }
    \label{tab:computation_21_gaussian}
    \begin{tabular}{cccccc}
    \toprule[1.5pt]
    \makecell[c]{21 cm model} &  \makecell[c]{ $t_{\rm{obs}}$\\ $\text{[orbit]}$ } & RMS [K] &  $A$ & $\nu_0$ & $\omega$ \\
    \midrule[1pt]
    Input  & --  & --& 0.20 & 75.00 & 6.00  \\
    \hline
    Recovered & 1 & 1.149 & $ 0.201_{-0.015}^{0.015} $ & $ 74.85_{-0.52}^{0.49}  $ & $ 6.11_{-0.55}^{0.61} $\\
    & 2 & 0.095 &  $ 0.200_{-0.010}^{0.010} $  & $ 75.04_{-0.36}^{0.34} $ & $ 5.93_{-0.39}^{0.38} $\\
    & 5 & 0.074 &  $ 0.2000_{-0.0067}^{0.0067} $  & $ 74.90_{-0.23}^{0.22} $ & $ 5.90_{-0.24}^{0.26} $\\
    & 10 & 0.041 &  $ 0.2020_{-0.0049}^{0.0048} $  & $ 74.91_{-0.17}^{0.16} $ & $ 6.04_{-0.18}^{0.20} $\\
    \midrule[1pt]
    Input  & --  & --& 0.155 & 75.00 & 10.00  \\
    \hline
    Recovered & 1 & 0.117 &  $ 0.170_{-0.026}^{0.032} $  & $ 75.3_{-1.4}^{1.2} $ & $ 10.2_{-1.5}^{1.6} $\\
    & 2 & 0.092 &  $ 0.160_{-0.020}^{0.017} $  & $ 75.27_{0.85}^{-0.94} $ & $ 9.6_{-1.0}^{1.1} $\\
    & 5 & 0.060 &  $ 0.162_{-0.012}^{0.014} $  & $ 75.11_{-0.67}^{0.62} $ & $ 10.22_{-0.70}^{0.76} $\\
    & 10 & 0.047 &  $ 0.1582_{-0.0082}^{0.0089} $  & $ 75.07_{-0.44}^{0.42} $ & $ 10.06_{-0.48}^{0.50} $\\
    \bottomrule[1.5pt]
    \end{tabular}
    \end{table*}

Next we consider the 21 cm signal extraction, from the mock data simulated with the foreground, the 21 cm signal, and the noise. 
We plot in Fig.\ref{fig:result} the five-term {\it LogPoly} model fitting results for the EDGES 21 cm model (left panel) and the two Gaussian 21 cm models (middle and right panels). We see that in all cases, the 21 cm model signals can be well recovered by fitting the mock observation data.

\begin{figure*}
\centering
\subfigure[]{
\includegraphics[width=0.8\columnwidth]{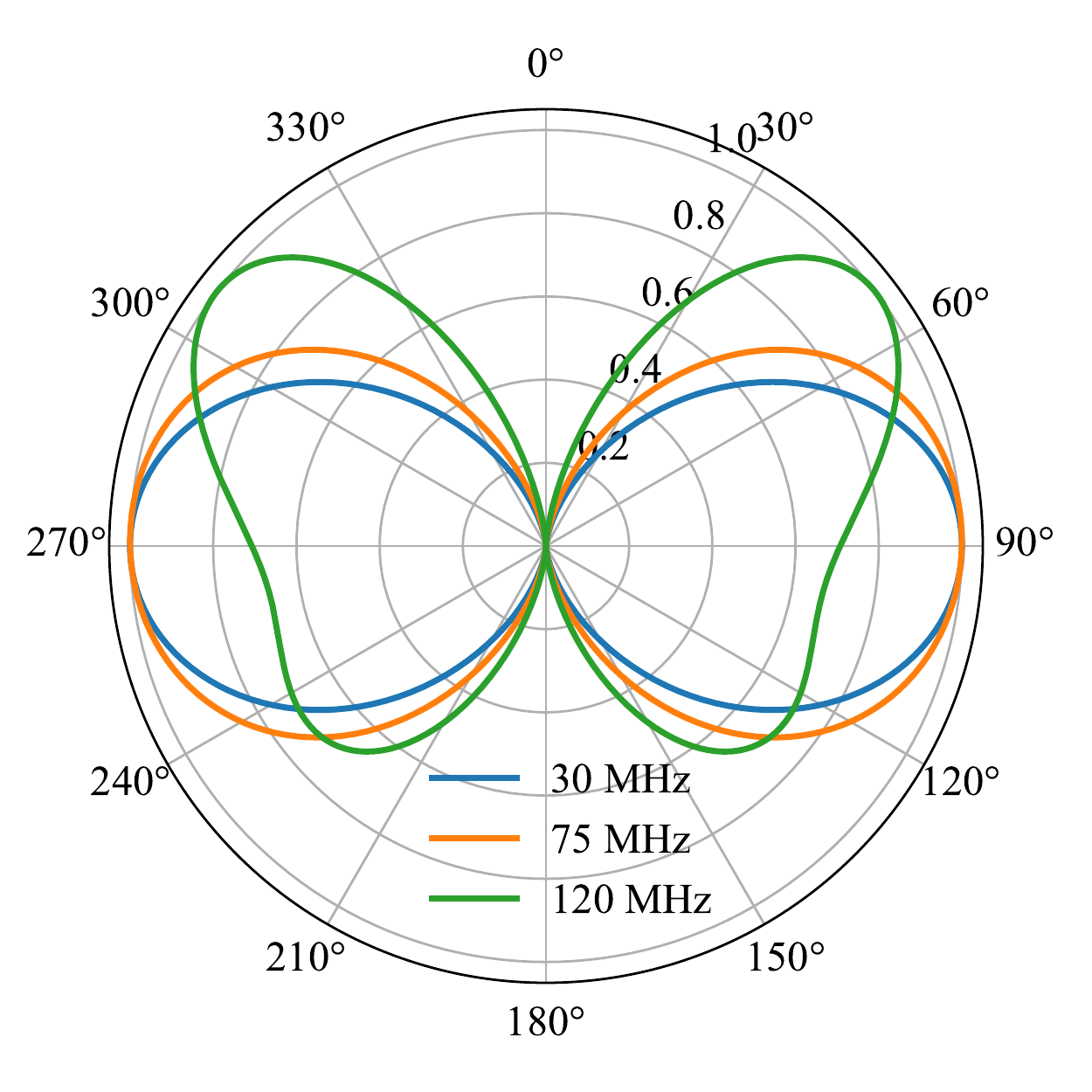}}
\subfigure[]{
\includegraphics[width=0.8\columnwidth]{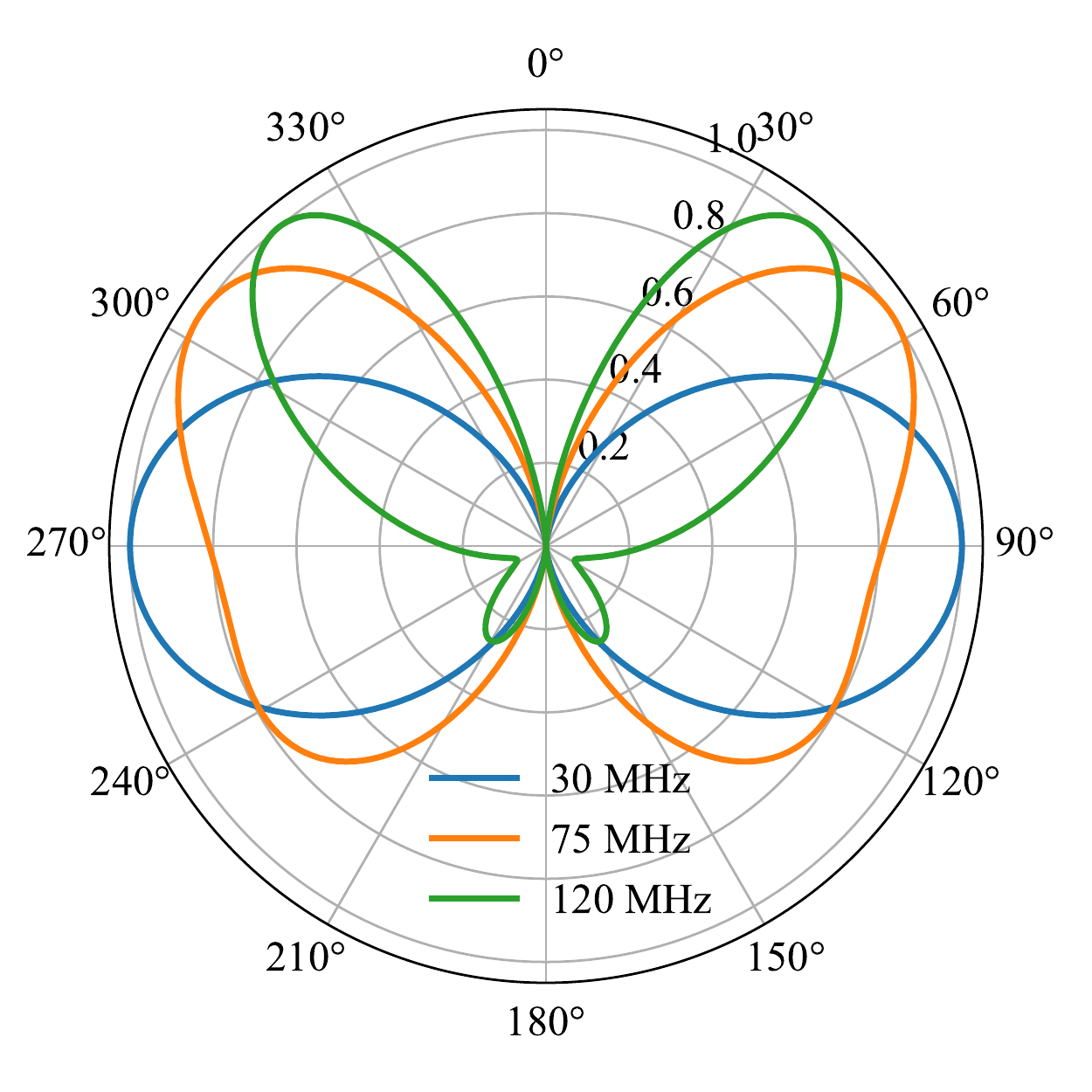}}\\
\subfigure[]{
\includegraphics[width=0.8\columnwidth]{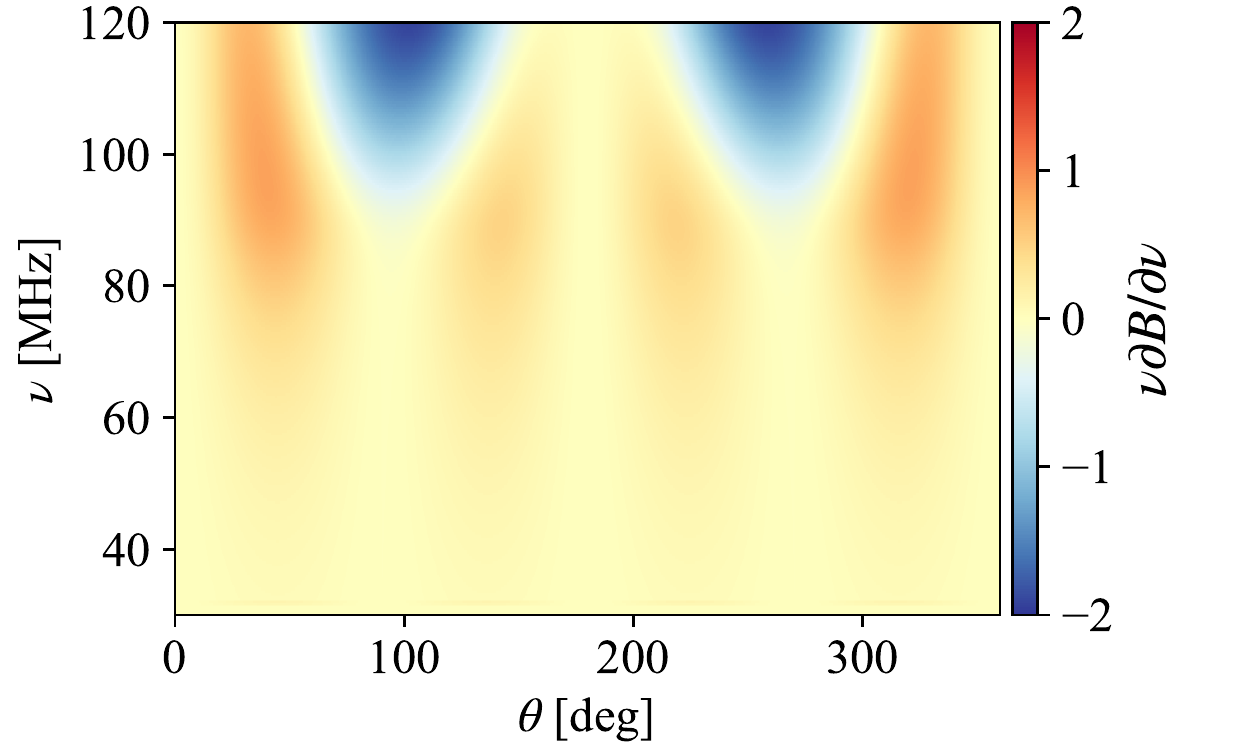}}
\subfigure[]{
\includegraphics[width=0.8\columnwidth]{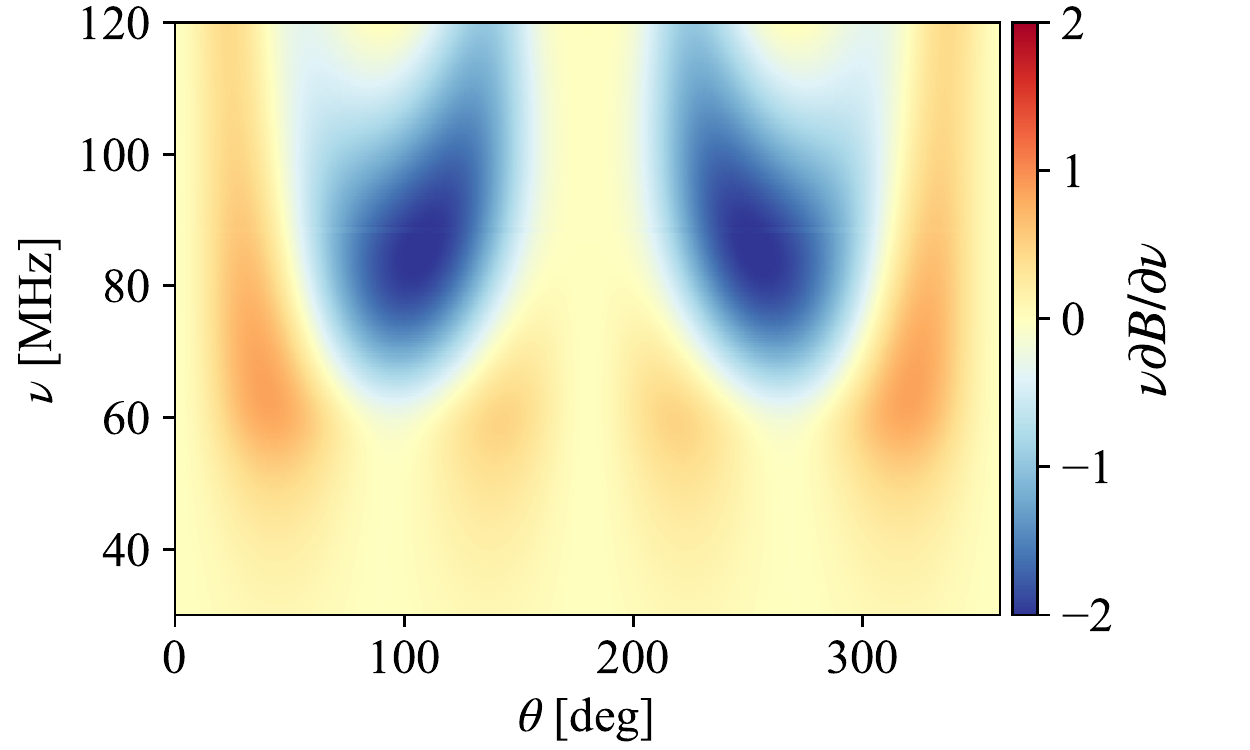}}
\caption{The cross-sections of beam profiles at a constant $\phi$ for the two villain models at three frequencies (top) and the corresponding frequency gradients at various $\theta$ and frequencies (bottom). The left panels (a, c) are for a disc-cone antenna with a 200 cm disc, and right panels (b, d) are for a disc-cone antenna with a 300 cm disc.}
\label{fig:2beams}
\end{figure*}

We also list the fitted model parameters for the EDGES 21 cm model in Table \ref{tab:computation_21}, and for the two Gaussian 21 cm models in Table \ref{tab:computation_21_gaussian} (the four rows correspond to the satellite completing 1, 2, 5 and 10 orbits, respectively). The thermal noise level drops with observation time as $t_{\rm{obs}}^{-1/2}$. For the frequency resolution of 0.4 MHz, the residual RMS induced by the thermal noise drops to $\sim 0.05 \K$ after 10 orbits, allowing the extraction of the 21 cm signal with good precision. The uncertainties of the parameters are estimated by Markov Chain Monte Carlo simulations. If the experiment is properly designed so that the error is dominated by thermal noise on this time scale, one should be able to discriminate between the EDGES model and the Gaussian models, and also between the two Gaussian models with different parameters introduced earlier. We also checked the results generated at different starting positions of precession for 10 effective orbits observation, the recovered 21 cm signals generally agree with each other within 0.01 K.

\subsection{Chromatic Beam Effect}

To explore how the chromatic beam affects the global spectrum measurement, we now consider some non-ideal cases where the beam is more frequency-dependent.  To be concrete,  we still consider disc-cone antennas but with different disc diameters ($D_3$ in Fig.~\ref{fig:antenna}).  However, we find that as long as the size of the antenna is smaller than the half wavelength, the beam is pretty similar to that of the short dipole and the frequency-dependence is weak. This is good news for the design of the antenna for the space experiment. However, for this illustration, we decide to employ two ``villain'' models, which are intentionally designed to have larger frequency variation of the beam.  These are realized by keeping the cone part unchanged, while increasing disc to a size comparable with the half wavelength. In the first  case, we increase the disc to 200 cm from the 55 cm in the fiducial model while keeping the cone unchanged. In the second case, the disc size is increased to 300 cm. One can obviously make antenna less frequency-dependent, but we use them as illustrative examples because they are derived from electromagnetic simulation of an antenna configuration. 

Fig.~\ref{fig:2beams} shows the beam profile of these two villain models in the top row, and the corresponding logarithmic frequency gradient of the beam in the bottom row. 
Comparing with Fig.~\ref{fig:beam_fid}, in these two cases the beams vary with frequency much more significantly, and the logarithmic frequency gradient are also much larger.  When convoluted with the spatial variations of the foreground, this may generate fake features in  the global spectrum. Furthermore, in the model with 
300 cm disc (right panels of Fig.~\ref{fig:2beams}),  an additional sidelobe appears. The presence of sidelobe could induce even larger features in the global spectrum from the sky variation, as we will find out below. 

To examine how this affects the measurement of the global spectrum, we simulate the observation of a total of 10 orbits with the foreground-only model without noise. Here for simplification we have used the data from the full orbit but ignored any possible RFI.

\begin{figure}[!thbp]
	\centering
    \includegraphics[width=0.95\columnwidth]{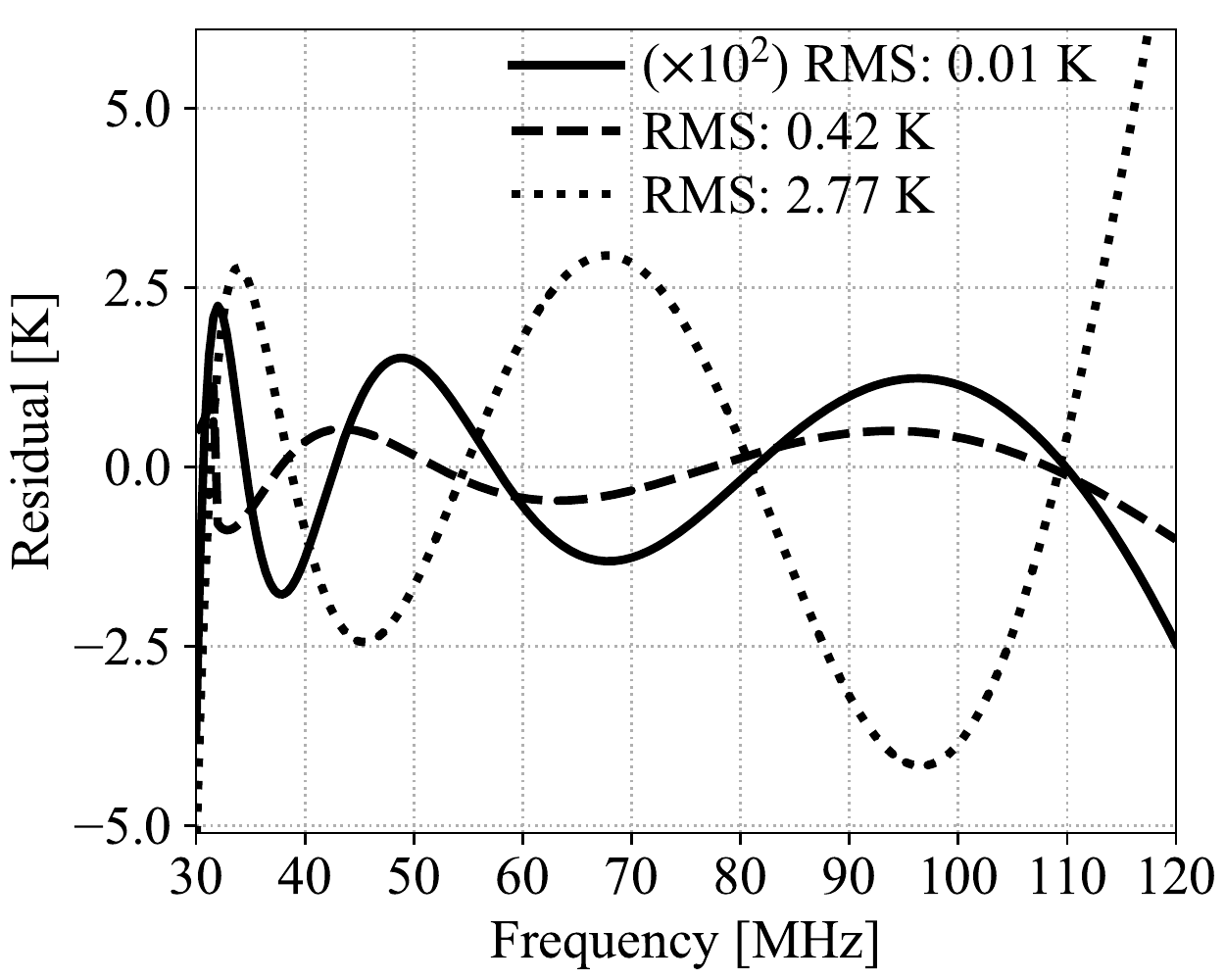}
    \caption{Residuals after fitting and subtracting a five-terms {\it LogPoly} model in frequency to the beam-weighted signal. 
    The solid, dashed, dotted lines correspond to results for $D=55 \ \cm$, 200 cm, and 300 cm antennas, respectively. 
    The residuals for the $D=55 \cm$ case is multiplied by a factor of 100 so that they can be plotted in the same figure.}
    \label{fig:gain_resi}
\end{figure}  

In Fig.~\ref{fig:gain_resi}, we plot the residuals after fitting and subtracting a five-term {\it LogPoly} model for our fiducial model, as well as for the two villain models. The residuals for the two villain models are about two orders of magnitude higher than
our fiducial model,  so we multiply a factor of 100 to the fiducial model residue to show them on the same plot. The RMS of the residuals of our fiducial model without adding the noise is about 10 mK.

As was shown in Fig.~\ref{fig:beam_fid}, $\nu \partial B/\partial\nu \sim 10^{-2}$ for our fiducial antenna beam model, and as shown in Fig.~\ref{fig:signal_A} the amplitude of the global spectrum varies at 10\% level. We may then expect the chromatic beam may introduce a deviation at the $10^{-3}$ level over the observation band, or $\sim$ 1 K for the $10^3 \sim 10^4 \K$ sky temperature of this band. However, the variation in the spectral shape or spectral index induced by the varying position/orientation is much smaller. As a result, this deviation varies smoothly over the frequencies, and so when we fit with the five-term {\it LogPoly} model, the bulk of it is removed. In the end, we obtain a residual with RMS of only 10 mK. This is still a very substantial and important source of error for the global spectrum measurement.

The two villain models
are both exaggerated bad cases, but these illustrate how the beam may vary with frequency over the observational band. Even for better designed cases, the frequency-gradient of the beam may have a pattern similar to them, though with smaller gradient values.  In particular, we note that the additional sidelobe may greatly increase the error. Thus, as shown in Fig.~\ref{fig:gain_resi}, the disc-cone antenna with a 300 cm disc which has an additional sidelobe in its beam has much larger RMS residuals than the one with a 200 cm disc. 

It is likely that the real antenna has a beam more like that of our fiducial model than that of the two villain models shown above. Nevertheless, there may be frequency-dependent deviations due to imperfect manufacture or additional components necessary for a satellite which will modify the antenna beam pattern. Such deviations could adversely affect the performance of the global spectrum measurement. It is critical to minimize such effects for the experiment to succeed.

\section{Conclusion and discussions}
\label{sec:conclusion}
In this work, we have presented the end-to-end simulation results for a global 21 cm spectrum experiment on lunar orbit operating in the 30 -- 120 MHz frequency band. By adopting the specific parameters of the DSL mission concept, we assessed the feasibility of obtaining the 21 cm global spectrum from a strong foreground. A variety of practical problems, such as the Moon blockage and antenna beam (including the chromatic beam effect) have been taken into account.
    
We first generate an input sky map with ISM free-free absorption effect, and make mock observation of the global spectrum. We find the absorption is significant 
at the lower frequencies, which affect the fitting of the foreground model. We simulate the observation at different positions on the lunar orbit. As the Moon blocks different parts of the sky at different positions, and the antenna orientation also changes with the position, there is variation in the measured sky signal. However, we find that though with the changing intensity in the received signal, the change in the spectrum shape (or spectral index) is much smaller. Adopting 
a five-term {\it LogPoly} model to fit the foreground, we find that the residual error drops to a level of  $\sim 0.05 \K$ within a short span of 10 orbits observation. The statistical noise would soon drop to a negligible level.   
    
We then consider systematic errors, such as the effect of chromatic beam. For our fiducial model with a disc-cone antenna with $D=55\  \rm{cm}$ for the disc, the beam chromaticity is actually very small. The variation of the gain is at the $10^{-4}$ to $10^{-3}$ level within the 30 -- 120 MHz range, and does not induce significant effect on the recovered 21 cm signal.  We also introduced two ``villain models'', which illustrated the effect of the chromatic beam with a poorly designed antenna, having a disc size of $D=200\ \rm{cm}$ or $D=300\ \rm{cm}$, exceeding the half wavelength. In this case the beam chromaticity can introduce much complexity in the result, especially if there is an additional sidelobe in the beam profile. In the real world, we expect the beam to be close to our fiducial case, perhaps with a slightly distorted profile and a little higher frequency-gradient due to the imperfections, but not large enough to affect the final result. In this study we have treated the Moon as a simple opaque sphere, and neglected the thermal emission of the Moon, and its reflection and diffraction of radio waves, which can be complicated due to its rugged terrain and varying composition. These problems will be further investigated in future studies.

Now it is feasible to conduct a new experiment towards the detection of the global 21 cm signal from cosmic dawn. The lunar orbit provides an unparalleled opportunity by allowing high precision global spectrum measurements. Our simulations show that the 21 cm global signal could be recovered very well on a lunar orbit.

\section*{Acknowledgements}
This work is supported by National SKA Program of China No. 2020SKA0110401,
the Chinese Academy of Sciences (CAS) Strategic Priority Research Program XDA15020200, the CAS Key Instrument Grant ZDKYYQ20200008, 
the National Natural Science Foundation of China (NSFC) grant 11973047, 11633004, and 
the CAS Frontier Science Key Project QYZDJ-SSW-SLH017.

\bibliography{main}{}
\bibliographystyle{aasjournal}



\end{CJK*}
\end{document}